\documentclass[12pt]{article}

\usepackage{booktabs}
\usepackage{amsmath}
\usepackage{amssymb}
\usepackage{hhline}
\usepackage[scale={.75,.75}]{geometry}
\usepackage{url}
\usepackage{caption}
\usepackage[numbers, sort&compress]{natbib}
\usepackage{epsfig}
\usepackage{multirow}
\usepackage{color}
\usepackage{verbatim}
\usepackage{nicefrac} 
\usepackage{upgreek}
\usepackage{bbm}
\usepackage{psfrag}
\usepackage{hyperref}

\newcommand{\bath}[1]{#1}

\newcommand{\q}{\textbf{q}}
\newcommand{\p}{\textbf{p}}
\renewcommand{\k}{\textbf{k}}
\renewcommand{\l}{\textbf{l}}

\newcommand{\Slash}[1]{\displaystyle{\not}{#1}}
\newcommand{\slashed}[1]{\displaystyle{\not}{#1}}

\newcommand{\fermionmass}[1]{\mathfrak{#1}}

\begin{document}
\date{\mbox{ }}

\title{ \vspace{-1cm}
%
\begin{flushright}
{\scriptsize \tt TUM-HEP-886/13, CAS-KITPC/ITP-367}  
\end{flushright}
{\bf \boldmath 
The Kinematics of Cosmic Reheating\footnote{
The present version of the article slightly deviates from the published version \cite{Drewes:2013iaa}. It includes a number of small corrections, which we summarised in a published Corrigendum.
}}
\\[8mm]}

\author{Marco Drewes$^{a}$, 
Jin U Kang$^{b}$\\  \\
{\normalsize \it$^a$ Physik Department T70, Technische Universit\"at M\"unchen,} \\{\normalsize \it James Franck Stra\ss e 1, D-85748 Garching, Germany}\\
{\normalsize \it$^b$ Department of Physics,  Kim Il Sung University,}\\
{\normalsize \it RyongNam Dong, TaeSong District, Pyongyang, DPR Korea}\\
}
\maketitle

\thispagestyle{empty}

\begin{abstract}
  \noindent
We calculate the relaxation rate of a scalar field in a plasma of other scalars 
and fermions with gauge interactions 
using thermal quantum field theory. It yields the rate of cosmic reheating and thereby determines the temperature of the ``hot big bang'' in inflationary cosmology. 
The total rate originates from various processes, including decays and inverse decays as well as Landau damping by scatterings. It involves quantum statistical effects and off-shell transport.
Its temperature dependence can be highly nontrivial, making it impossible to express the reheating temperature in terms of the model parameters in a simple way.
We pay special attention to the temperature dependence of the phase space due to the modified dispersion relations in the plasma.  
We find that it can have a drastic effect on the efficiency of perturbative reheating, which depends on the way particles in the primordial plasma interact. 
For some interactions thermal masses can effectively close the phase space for the dominant dissipative processes and thereby impose an upper bound on the reheating temperature. In other cases they open up new channels of dissipation, hence increase the reheating temperature.
At high temperatures we find that the universe can even be heated through couplings to fermions, which are often assumed to be negligible due to Pauli-blocking.
These effects may also be relevant for baryogenesis, dark matter production, the fate of moduli and in scenarios of warm inflation. 
\end{abstract}

\newpage\tableofcontents
\newpage
\section{Introduction
}\label{intro}
In this article we calculate the relaxation rate $\Gamma$ of a scalar field $\phi$ in a plasma of other scalars and fermions with gauge interactions using thermal quantum field theory. We study in detail which effect screening and the modified dispersion relations of quasiparticles in a hot plasma have on $\Gamma$.  
If $\phi$ is identified with the {\it inflaton} field driving cosmic inflation, then $\Gamma$ gives the rate at which the universe is heated up after inflation and thus determines the temperature of the {\it hot big bang} (``reheating'').
However, our results can also be applied to estimate the dissipation rate during inflation itself, which is of great interest in models of warm inflation.
Furthermore, $\phi$ need not be identified with the inflaton and may also be interpreted as the order parameter in a phase transition in the early universe. 
Finally, the kinematic effects we study can also be relevant for other transport phenomena that involve particle production or dissipation, such as baryogenesis or dark matter production. In this article we focus on cosmic reheating and do not discuss other applications in detail; a short list and references are given in the conclusions in section \ref{Conclusions}.

\subsection{The big bang temperature}
Many properties of the cosmos we observe today can be understood as the result of processes that occurred during the early, high temperature phase of its history \cite{Kolb:1990vq}. 
This makes the maximal temperature
in the early universe a crucial parameter in physical cosmology \cite{Buchmuller:2012tv}. 
An important probe of the thermal history of the universe is given by the abundance of thermal relics.
These include the light elements that we observe in the intergalactic medium. They were created in thermonuclear reactions in the primordial plasma, known as big bang nucleosynthesis (BBN).
The good agreement between theoretical BBN calculations and observation allows to conclude that the universe in the past had a temperature $T_{\rm BBN}$ of at least a few MeV. 

Observationally, only little is known about the thermal history of the universe beyond that.
However, there are good reasons to believe that it has been exposed to much higher temperatures.
If the observed dark matter (DM) 
is a thermally produced {\it weakly interacting massive particle}, then temperatures not too much below that particle's mass are required for efficient thermal production.
Also other DM scenarios often require temperatures $T\gg T_{\rm BBN}$, for instance sterile neutrinos \cite{Dodelson:1993je,Asaka:2005pn} (see \cite{Drewes:2013gca} for a recent review) or axions (see \cite{Kim:2008hd} for reviews). 
The observed baryon asymmetry of the universe (see \cite{Canetti:2012zc} for a recent review) provides another hint for large temperatures.
If the baryon number violation in the Standard Model of particle physics (SM) is the relevant source of the baryon asymmetry \cite{Kuzmin:1985mm}, 
then the universe must have been exposed to temperatures $T>T_{\rm EW}\sim 140$ GeV unless an unknown mechanism prevented the SM interactions from driving the plasma to thermal equilibrium.\footnote{This conclusion is based on the fact that baryon number violating processes in the SM are strongly suppressed in a thermal plasma for $T<T_{\rm EW}$, while the CP-violation is small for $T\sim T_{\rm EW}$, as e.g. reviewed in \cite{Canetti:2012zc}. At lower temperatures the CP-violation is larger \cite{Brauner:2011vb}, but baryon number violation can only be realised in the SM if the universe is exposed to a non-standard thermal history involving large deviations from equilibrium \cite{GarciaBellido:1999sv}.}    
Many models of baryogenesis require a temperature that is even much higher than that, including GUT-baryogenesis \cite{Kolb:1983ni}, thermal leptogenesis \cite{Fukugita:1986hr}, baryogenesis from neutrino oscillations \cite{Akhmedov:1998qx,Asaka:2005pn,Drewes:2013gca} or many variants of Affleck-Dine baryogenesis \cite{Affleck:1984fy}. 
Hence, any upper bound on the temperature in the universe allows to rule out or constrain models of particle physics beyond the standard model (SM).  
A well-known example is the conflict between supersymmetry and thermal leptogenesis due to the gravitino problem \cite{Pagels:1981ke}.
A large temperature can also be related to the fate of moduli \cite{Yokoyama:2006wt} and 
be of great importance for the possible decompactification of extra dimensions \cite{Buchmuller:2004xr}. 

\subsection{The origin of the hot big bang}
It is widely believed that the {\it hot big bang} initial conditions of the radiation dominated universe were created during \textit{reheating} after a period of accelerated cosmic expansion, i.e. {\it cosmic inflation} \cite{Starobinsky:1980te,Guth:1980zm,Linde:1981mu}.
Inflation made the universe homogeneous and flat, while creating the seeds for later structure formation from quantum fluctuations \cite{Mukhanov:1981xt}.
The accelerated expansion can easily be explained if one assumes that the energy density of the universe was dominated by the potential energy $V(\phi)$ of a scalar field $\phi$, 
the inflaton,\footnote{For simplicity, we consider a single field model of inflation. However, the considerations in section \ref{RehMech} apply to any model of inflation that includes perturbative reheating.} which leads to a negative equation of state while $\phi$ is slowly rolling towards the potential minimum.
At the end of the inflationary epoch, the universe was cold and empty, with all energy stored in the zero mode of $\phi$. 
All other particles were then created and the universe was (re)heated \cite{Dolgov:1982th,Traschen:1990sw,Kofman:1994rk,Boyanovsky:1994me} by the dissipation during coherent oscillations of $\phi$ around its potential minimum.
This dissipation can be driven by perturbative processes, such as the decay of individual inflaton quanta into lighter particles or inelastic scatterings in the plasma. These are the processes we study in this article. 
In the simplest scenarios these are the only processes at work, and 
our methods allow to completely describe the reheating process. 

However, the rapidly oscillating $\langle\phi\rangle$-condensate can act as a non-adiabatic background to which the other fields couple. 
Because of this, early stage of reheating, sometimes referred to as preheating, may have involved complicated nonequilibrium processes such as nonperturbative particle production \cite{ES1939,Traschen:1990sw,Kofman:1994rk,Boyanovsky:1994me}, turbulence \cite{Micha:2004bv} and non-thermal fixed points \cite{Berges:2008wm} or instabilities \cite{GarciaBellido:1997wm,Felder:2000hj}, including explosive particle production by parametric resonance \cite{Traschen:1990sw,Kofman:1994rk,Khlebnikov:1996mc,Berges:2002cz}. 
We are not concerned with these mechanisms in this article;
their quantitative description usually requires expensive numerical simulations. 
Some of the relevant effects have recently been discussed in detail in \cite{Mukaida:2012qn,Mukaida:2012bz}.
If they occur, they mainly affect the early preheating phase.
The late phase of reheating, which determined the temperature at the onset of the radiation dominated epoch, in many scenarios can be treated perturbatively.
In addition, perturbative processes are also of great interest during preheating if they allow the produced particles to decay efficiently within one $\phi$-oscillation \cite{Felder:1998vq,Micha:2004bv,GarciaBellido:2008ab}. This can delay or completely avoid the parametric resonance.
Then nonperturbative production ceases while the energy budget is still $\langle\phi\rangle$-dominated and the universe is essentially reheated by perturbative processes. 
Thus, a quantitative understanding of perturbative processes involving individual inflaton quanta is crucial to understand the hot big bang initial conditions, in particular the {\it reheating temperature} $T_R$.

It is common to make a ``naive'' estimate for the reheating temperature  by assuming that a) the inflaton converts its entire energy into radiation instantaneously in the moment when $\Gamma=H$, where $H$ is the Hubble rate, and
b) this radiation comes into thermal equilibrium instantaneously. 
Due to condition b) we can replace $H=\sqrt{8\pi^3 g_*/90}\phantom{i} T^2/M_P$, where $M_P$ is the Planck mass and $g_*$ the number of relativistic degrees of freedom in the plasma.
By setting $H=\Gamma$ we find
\begin{equation}\label{naive}
T_R\equiv\left(\frac{90}{8\pi^3 g_*}\right)^{1/4}\sqrt{\Gamma M_P}.
\end{equation}
If we require as a third condition c) that the radiation- and matter-density is negligible prior to the moment $\Gamma=H$,
then $\Gamma$ can be identified with the vacuum decay rate $\Gamma|_{T=0}$ of $\phi$-particles. In this case the RHS of (\ref{naive}) is independent of $T$ and the couplings of other fields amongst each other.

In reality condition a) is of course not fulfilled: The inflaton dissipates its energy at a finite rate $\Gamma$, and the reheating process takes place over a time interval $\sim 1/\Gamma$. 
In spite of this, (\ref{naive}) provides an estimate for the reheating temperature (defined as the temperature at the onset of the radiation dominated era) if 
conditions b) and c) are fulfilled. 
This is, however, in general not the highest temperature in the early universe, which we refer to as $T_{\rm MAX}$.
At first sight one could argue that (\ref{naive}) provides an upper bound on the temperature in the early universe: As the radiation cools down due to the universe's expansion during the time span $\sim 1/\Gamma$, one might expect that the temperature is always lower than (\ref{naive}).
This argument is incorrect because condition a) is not fulfilled, i.e. dissipation already starts before the moment
$\Gamma=H$. Even though the fraction by which the inflaton's energy density $\rho_\phi$ is reduced per
Hubble time is very small prior to $\Gamma=H$, the amount of
energy released into radiation is larger than at $\Gamma=H$ and later
times because of the larger $\langle\phi\rangle$-amplitude (and hence larger $\rho_\phi$). 
Therefore the maximal temperature is reached well before $\Gamma=H$ and larger than $T_R$ \cite{Giudice:2000ex}. This has the linguistically curious consequence that $T$ actually decreases during most of the reheating process.
It can be estimated as \cite{Giudice:2000ex}
\begin{equation}
T_{\rm MAX}\simeq 0.7 T_R^{1/2}\left(\frac{V_I}{g_*}\right)^{1/8}\label{TmaxNaive}.
\end{equation}
Here $V_I$ is $V(\phi)$ at initial time.
The estimates (\ref{naive}) and (\ref{TmaxNaive}) are indeed rather accurate as long as conditions b) and c) are fulfilled. 
In this paper we are mostly concerned with the consequences of relaxing condition c). 
Even if there is no phase of preheating prior to perturbative reheating, a radiation bath is formed by the decay products shortly after reheating commences.
In order to determine the time evolution of $T$, its maximal value and its value at the onset of the radiation dominated era, it is crucial to take into consideration the effect this medium has on $\Gamma$ during reheating \cite{Kolb:2003ke,Yokoyama:2005dv,Drewes:2010pf,Mukaida:2012bz,Harigaya:2013vwa,Drewes:2014pfa}.
Indeed, analytic and numerical estimates consistently show that it can change $T_{\rm MAX}$ by orders of magnitude \cite{Drewes:2014pfa}.

The presence of the primordial plasma affects the rate $\Gamma$ in various ways: Quantum statistical factors can enhance or suppress individual channels due to Bose enhancement or Pauli blocking. The modified dispersion relations of particles in the hot plasma make the phase space temperature-dependent and new channels of dissipation can open up in addition to the decay of $\phi$-quanta, such as Landau damping.  
As we will discuss in the following, this may greatly enhance or suppress $\Gamma$ and have a strong effect on both, the maximal temperature the universe has been exposed to and the temperature at the beginning of the radiation dominated epoch.

\subsection{The role of ``thermal masses''}
In most models of inflation the inflaton mass $m$ is larger than the masses of known particles.\footnote{Exceptions from this include Higgs inflation \cite{Bezrukov:2007ep}, for which reheating has been discussed in \cite{Bezrukov:2008ut,GarciaBellido:2008ab}, and the model introduced in \cite{Shaposhnikov:2006xi}.} In any case, the SM is in the symmetric phase for $T>T_{\rm EW}$. In vacuum the only perturbative mechanism by which $\phi$ dissipates its energy into other degrees of freedom is the decay of heavy $\phi$ quanta into lighter states. The particles in the universe today are the end products of the subsequent decay chain.
However, once $\phi$ has lost a significant fraction of its energy, the universe is filled with a medium formed by the hot plasma of its decay products. For the moment we collectively refer to these as $\mathcal{X}_i$, without specification of the spin, charges or other quantum numbers of these fields. The $\mathcal{X}_i$ represent all degrees of freedom other than $\phi$, including the SM fields. 
The inflaton relaxation time, characteristic for the duration of reheating, is given by $\tau\sim 1/\Gamma$, where $\Gamma$ is the $\phi$ relaxation rate in the medium.
The coupling of $\phi$ to other fields should be very small in order not to spoil the flatness of its potential required for slow roll inflation \cite{Finelli:2009bs}. 
This implies that $\tau$ is much longer than the typical time scale associated with the relaxation to thermal equilibrium for any other field in the plasma.\footnote{We do not make any other assumptions about the strength of the inflaton coupling. It may indeed be a non-trivial requirement that it is large enough to allow for efficient reheating and small enough not to spoil the flatness of the inflaton potential. We do not address this issue here and express all contributions to $\Gamma$ in units of their zero temperature value, in which case its dependence on the inflaton coupling cancels at leading order.} 
This suggests that it is reasonable to approximate perturbative reheating as an adiabatic process, during which the $\mathcal{X}_i$ are close to kinetic equilibrium at any time\footnote{In principle the thermalisation process is not trivial, see \cite{Enqvist:1990dp,Harigaya:2013vwa} for some recent discussion.
One problem that has been observed in bosonic toy models is that most energy is stored in the infrared modes (especially during preheating). This can significantly prolong thermalisation because scatterings amongst low momentum particles fail to populate high momentum modes. However, in more realistic models involving fermions equilibration can occur considerably faster \cite{Berges:2010zv}. 
}. 

It is well-known that in quantum field theory the spectrum of propagating states (``quasiparticles'') and their dispersion relations are modified in a medium due to screening and collective phenomena.
In the following we assume for simplicity that the $\mathcal{X}_i$ are in kinetic equilibrium and can be characterised by an effective temperature $T$ \footnote{We make this assumption for computational simplicity; the arguments brought forward in section \ref{RehMech} remain valid even if the background is far from equilibrium \cite{Drewes:2012qw}. }. 
Then thermal corrections to the dispersion relations typically scale as $\sim\alpha_i T$. Here $\alpha_i$ are dimensionless numbers that characterise the strength of a typical interaction amongst the $\mathcal{X}_i$. They can e.g. represent gauge or Yukawa coupling constants. In the simplest case, the system can be pictured as a gas of quasiparticles with momentum independent thermal masses $M_i\sim \alpha_i T$.

$\phi$ should have very weak interactions to preserve the flatness of its potential, but the $\alpha_i$ at $T\gg T_{\rm EW}$ can be much bigger. They may be of a similar size as the SM gauge couplings or of order one. Thus, the thermal masses of the decay products $\mathcal{X}_i$ grow much faster with temperature than the effective $\phi$ mass.
In \cite{Kolb:2003ke} it has been concluded that due to this effect, the inflaton decay becomes kinematically forbidden at some critical temperature $T_c$, when the sum of the effective masses of the decay products exceeds the inflaton mass,
\begin{equation}
M(T_c)=\sum_{{\rm daughter} \ {\rm particles}\ i}M_i(T_c) \label{TcDef}
\end{equation}
As a result, the universe could never be heated up to temperatures larger than $T_c$.
Due to the very small couplings of $\phi$, thermal corrections to the $\phi$ dispersion relations are tiny and the LHS of (\ref{TcDef}) can essentially be identified with the vacuum mass $m$.\footnote{
The generation of an effective mass in the radiation dominated era has e.g. been studied in \cite{Kawasaki:2011zi}.
} Thus $T_c$, defined by (\ref{TcDef}), in good approximation does not depend on the inflaton coupling.
If $T_c$ imposes an upper bound on the reheating temperature, this would have far reaching consequences as it would allow to constrain the temperature in the early universe without knowledge of the inflaton couplings.

In \cite{Yokoyama:2005dv} it was shown for a simple toy model that $\Gamma$ is non-vanishing in the limit $T\gg T_c$. However, the analysis does not take into account the fact that $\Gamma$ may be vanishing for intermediate temperatures $T\sim T_c$; in this case $T_c$ would still pose an upper bound on $T$, as the universe could never pass this ``forbidden regime'' of temperatures. We indeed find that in the toy model used in \cite{Yokoyama:2005dv} this is the case 
because $\Gamma$ is  strongly suppressed at $T\sim T_c$. 
The problem was studied in more detail in \cite{Drewes:2010pf}, where a physical interpretation in terms of scatterings was given. 
In the present work we extend that study and perform a detailed analysis of the kinematics in a hot plasma and its effect on $\Gamma$. We show that the suppression of $\Gamma$ at $T_c$ due to thermal masses only occurs in special cases; in general they do not impose an upper limit on the reheating temperature. We also find that at high temperatures the dissipation into fermions can give a significant contribution to $\Gamma$.

\subsection{Overview over this article}
In section \ref{RehMech} we argue on general grounds that the kinematics of reheating can often not be described correctly by simply replacing vacuum masses of particles by thermal masses. 
In sections \ref{example1}-\ref{yukawasection} we illustrate this by calculating the inflaton relaxation rate $\Gamma$ in a model of chaotic inflation for different interactions. We first study the dissipation into bosons. We chose three different examples to represent cases where thermal masses block the inflaton decay (section \ref{example1}), do not block the decay (section \ref{example2}), or Landau damping drives reheating (section \ref{QuarticInteraction}).
In section \ref{yukawasection} we study the inflaton coupling to fermions with gauge interactions. At low temperatures the decay into fermions is suppressed due to Pauli blocking before thermal masses become relevant, but at large temperatures $\phi$ can efficiently dump its energy into fermions via Landau damping. 
In section \ref{Conclusions} we discuss our results and conclude. The reader who is not interested in technical details may skip sections \ref{elements}-\ref{yukawasection} and read sections \ref{intro}, \ref{RehMech} and \ref{Conclusions} as a short letter.

\section{Summary: The kinematics of cosmic reheating}\label{RehMech}
Due to the medium corrections to dispersion relations the phase space in a hot plasma is temperature dependent. This can have a strong effect on the efficiency of reheating. 
When the primordial plasma is modelled as a gas of quasiparticles, 
it is often assumed that  their kinematics can be described by simply replacing vacuum masses of particles by momentum independent thermal masses $\sim \alpha_i T$.
In the context of reheating, such approximation may fail even at the qualitative level, as the following effects need to be taken into account. 
\begin{itemize} 
\item {\bf Particle production} - The inflaton field $\langle\phi\rangle$ oscillates with a frequency $\sim m$, which is higher than the mass of most other particles. In such a rapidly time-varying background it is not possible to define an appropriate vacuum state and particles \cite{ES1939}. In such a background dissipation is not due to perturbative processes, but to particle production from the background. 
In this regime the coupling to the $\langle\phi\rangle$-background can give rise to the explosive particle production when it dominates during preheating, and arguments based on single particle kinematics certainly fail. A quantitative treatment of this preheating process goes beyond the scope of this work.

\item {\bf Quasiparticle description} - When the produced particles start making up some fraction of the universe's energy content, they form a thermal background, to which we assign an effective temperature $T$. 
The existence of well-defined quasiparticles in this plasma is related to the requirement that all dressed spectral densities [to be defined later in (\ref{spectraldensitydefinition})] as a function of energy $\omega$ feature narrow peaks at energies $\omega=\Omega_i$, which define the dispersion relations of quasiparticles.  
These receive three contributions: \textit{intrinsic (vacuum) masses} $m_i$, a time-dependent mass due to the coupling to the oscillating mean field $\langle\phi\rangle$ and \textit{thermal masses} 
due to forward scatterings in the plasma. 
A (quasi)particle description is appropriate if the typical energy $\omega$ of a $\mathcal{X}_i$-mode does not change considerably during a scattering process \cite{Drewes:2012qw}. This is the case if either  the inflaton's oscillation frequency is small compared to $\omega$ or
the contribution to the particles' effective masses from
 the slowly changing 
thermal background 
dominates over that from the rapidly oscillating $\langle\phi\rangle$-condensate. 
This is possible even if the energy stored in $\langle\phi\rangle$ exceeds the energy of the primordial plasma because the inflaton couples weakly: 
If $g$ is the inflaton coupling and $\alpha_i$ a typical coupling within the plasma (with $\alpha_i\gg g$), then thermal contributions $\sim\alpha_i T$ can dominate over $g\langle\phi\rangle$ even if $\langle\phi\rangle\gg T$. 
In this case reheating can be described in terms of perturbative processes involving individual quasiparticles and phase space arguments apply. This is the regime we focus on in this work.
Since the $\langle\phi\rangle$-amplitude decreases while $T$ increases in the course of reheating, it becomes an increasingly good approximation at later times. 

\item  {\bf Landau damping} - In the regime where a quasiparticle description is appropriate, large thermal masses may kinematically block the inflaton decay.
However, decay is not the only channel by which $\phi$ can dissipate energy into the plasma. 
$\phi$-quanta can engage in scatterings with particles from the bath, leading to a dissipation mechanism similar to Landau damping \cite{Weldon:1983jn}. At low temperatures, this mechanism is inefficient because the concentration of scattering partners is too low. At high temperatures it becomes the dominant channel of dissipation because 
processes involving several interactions in a given time unit become equally likely as those involving one or few when the density is high. Landau damping is the main reheating mechanism in the model discussed in section \ref{QuarticInteraction}, but also appears in sections \ref{example1}, \ref{example2} and \ref{yukawasection}.

\item  {\bf Quasiparticle dispersion relations} - The dispersion relations for quasiparticles in most cases cannot be obtained  by simply replacing bare masses with momentum independent thermal masses. In general, the quasiparticle energy $\Omega_i$ is a complicated function of momentum $\textbf{p}$. 
In spite of that, many transport phenomena can effectively be described by thermal masses $M_i\sim\alpha_i T$. The reason is that for hard momenta $\p\sim T$ the dispersion relations usually asymptotically approach a form 
$\Omega_i\simeq(\p^2+M_i^2)^{1/2}$. 
Since most particles in a plasma have momenta $\sim T$, a description in terms of momentum independent thermal masses is appropriate for most microscopic processes. However, when the inflaton decays, the decay products have typical momenta $\lesssim m/2$ (and not $\sim T$). 
For $m<T$ these are "soft" compared to typical energies $\sim T$ in the plasma. The dispersion $\Omega_i$ relations for soft modes strongly depend on the details of the interactions; they can be complicated functions of momentum (see sections \ref{example2} and \ref{yukawasection}). It is not necessarily true that these modes become "thermally heavy" and kinematically block the inflaton decay, see section \ref{example2}. 

\item {\bf Collective excitations} - It is well-known \cite{LeB} that for soft momenta there can be collective excitations in a plasma (holes/plasminos, plasmons, etc.) that behave like quasiparticles
in addition to the screened versions of particles known from vacuum \cite{Klimov:1981ka,Weldon:1982bn,Drewes:2013bfa}, see \cite{Hidaka:2011rz,Nakkagawa:2011ci} for a recent discussion.
Decays into these and scatterings with them may give a contribution to $\Gamma$ \cite{Enqvist:2004pr,Kiessig:2010pr}, as we see in section \ref{yukawasection}.

\item {\bf Off-shell transport} - In quantum mechanics processes involving virtual (off-shell) particles contribute to transition amplitudes. These contribute to the energy transfer between the inflaton and other fields even if some of the intermediate states are too heavy to be produced as real (quasi)particles. In our setup the relevance of off-shell processes can be parametrised in terms of the thermal $\mathcal{X}_i$-widths $\Gamma_i$. 
Though in principle suppressed by $\Gamma_i/\Omega_i$, the off-shell transport can give a sizable contribution to $\Gamma$ due to collinear amplification or large occupation numbers 
\cite{Yokoyama:2005dv,Garbrecht:2008cb,Anisimov:2010gy,Drewes:2010pf,Hamaguchi:2011jy}.
\end{itemize}
These considerations 
are very general.
In the following sections we illustrate them in a simple model of chaotic inflation. 
The reader who is not interested in technical details may skip them and read sections \ref{intro}, \ref{RehMech} and \ref{Conclusions} as a short letter.
\section{Elements of thermal quantum field theory}\label{elements}
The crucial quantity that determines how efficiently $\phi$ can reheat the universe is the relaxation rate $\Gamma$. 
It determines how much energy $\phi$ dissipates into the primordial plasma per unit time.
The ``in-out formalism'' and S-matrix, commonly used in quantum field theory in vacuum, do not provide an appropriate tool to describe nonequilibrium phenomena in a dense plasma. The reason is that a nonequilibrium process is an initial value problem, in which the final state is not known a priori and memory effects can be important. Furthermore, the definition of asymptotic states is ambiguous in the omnipresent plasma when the density is large enough that particles always feel the presence of their neighbours.  
A consistent treatment is possible in the framework of nonequilibrium quantum field theory, where all properties of the system can be expressed in terms of correlation functions of quantum fields.\footnote{For a review see \cite{Berges:2004yj}.} The approach we use in the following is known as Schwinger-Keldysh formalism \cite{Schwinger:1960qe}, but sometimes also referred to as ``closed time path'' or ``in-in formalism''.
We use the notation of \cite{Anisimov:2008dz,Drewes:2010pf,Anisimov:2010aq}.

In the Schwinger-Keldysh formalism, the gain and loss rates $\Gamma^<_\q$ and $\Gamma^>_\q$ for the  $\phi$-mode $\q$ are related to self-energies $\Pi^<(x_1,x_2)$ and $\Pi^>(x_1,x_2)$.
Inflaton couplings that are linear in $\phi$ 
can be expressed as $\phi\mathcal{O}[\mathcal{X}_i]$, where $\mathcal{O}[\mathcal{X}_i]$ represents operators that are composed of fields $\mathcal{X}_i$ other than $\phi$. 
In the following sections we will use the model Lagrangian
\begin{eqnarray}\label{L}
\mathcal{L}&=& 
\frac{1}{2}\partial_{\mu}\phi\partial^{\mu}\phi-\frac{1}{2}m^{2}\phi^{2}
+\bar{\Psi}\left(i\Slash{\partial}-\fermionmass{m}\right)\Psi
+ \sum_{i=1}^2 
\left(
\frac{1}{2}\partial_{\mu}\chi_{i}\partial^{\mu}\chi_{i}
-\frac{1}{2}m_{i}^{2}\chi_{i}^{2}
\right)
\nonumber\\
&&-g\phi\chi_{1}\chi_{2}-Y\phi\bar{\Psi}\Psi-\sum_{i=1}^{2}
\frac{h_{i}}{4!}\phi\chi_{i}^{3}+\mathcal{L}_{\mathcal{X}}
.\end{eqnarray}
The $\chi_i$ are scalar and $\Psi$ is a fermionic field to which $\phi$ couples. $\mathcal{L}_{\mathcal{X}}$ contains all other degrees of freedom (including the SM fields) and their couplings to the $\chi_i$ and $\Psi$. 
In this section we do not specify the interactions in $\mathcal{L}_{\mathcal{X}}$;
as before, we symbolically characterise their strength by dimensionless numbers $\alpha_i$.  It is assumed that $m\gg m_i, \fermionmass{m}$ and $g/m, Y, h_i\ll \alpha_i$.
For (\ref{L}) the operator $\mathcal{O}[\mathcal{X}_i]$ is identified with
\begin{equation}
\mathcal{O}[x]=g\chi_{1}(x)\chi_{2}(x)
+Y\bar{\Psi}(x)\Psi(x)+\sum_{i=1}^{2}
\frac{h_{i}}{4!}\chi_{i}^{3}(x).
\end{equation}
To leading order in the tiny inflaton couplings the self-energies can be calculated as
\begin{equation}
\Pi^>(x_1,x_2)=\langle\mathcal{O}(x_1)\mathcal{O}(x_2)\rangle,\ \Pi^<(x_1,x_2)=\langle\mathcal{O}(x_2)\mathcal{O}(x_1)\rangle.
\end{equation}
The average $\langle\ldots\rangle$ is defined in the usual way as $\langle\mathcal{A}\rangle={\rm Tr}(\varrho \mathcal{A})$, where $\varrho$ is the density matrix of the thermodynamic ensemble under consideration. It includes the usual quantum average as well as a statistical average over initial conditions.
It is convenient to define the {\it spectral self-energy},
\begin{equation}\label{PiDef}
\Pi^-(x_1,x_2)=\Pi^>(x_1,x_2)-\Pi^<(x_1,x_2)
,\end{equation}
which can be related to the usual retarded self-energy by $\Pi^R(x_1,x_2)=\theta(t_1-t_2)\Pi^-(x_1,x_2)$. 
$\Pi^-(x_1,x_2)=\Pi^>(x_1,x_2)-\Pi^<(x_1,x_2)$ is sometimes also referred to as {\it dissipative self-energy} because it determines the total relaxation rate for $\phi$, including gain- and loss terms \cite{Weldon:1983jn}, cf. (\ref{GammaFormula}).
In contrast to that, the combination $[\Pi^>(x_1,x_2)+\Pi^<(x_1,x_2)]{\rm sign}(x_1-x_2)$ is called {\it dispersive self-energy}. In momentum space the dissipative and dispersive self-energy proportional to the imaginary and real part of the retarded self-energy, cf. (\ref{ImPiRvsPiMinus}).

\begin{figure}
  \centering
    \includegraphics[width=6cm]{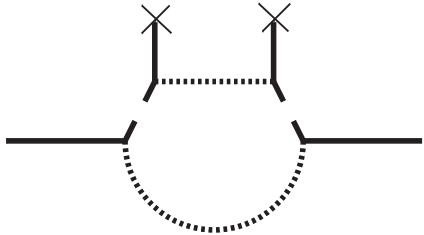}
    \caption{An example for a contribution to $\Pi^-$ that involves the mean field; crosses represent the mean field $\langle\phi\rangle$, lines represent propagators for $\phi$ (solid), $\chi_1$ (dashed) and $\chi_2$ (dotted).\label{meanfieldfigure}}
\end{figure}

In the following we will for simplicity always assume that all fields other than $\phi$ are in thermal equilibrium with temperature $T$ (they relax on time scales $1/\Gamma_i\ll 1/\Gamma$, while we are interested in the time scale $1/\Gamma$, which is characteristic for the duration of reheating\footnote{In reality thermalisation is a highly non-trivial process, see e.g. \cite{Allahverdi:2011aj,Kurkela:2011ti}. However, most of the following can be generalised in a straightforward way as long as $\Gamma\ll m$  \cite{Drewes:2012qw}.}.
Due to the homogeneity and isotropy of the universe,  $\Pi^-$ depends only on the relative spacial coordinate $|\textbf{x}_1-\textbf{x}_2|$.
To leading order in the tiny couplings $g, h_i$ and $Y$, the self-energy $\Pi^-$ is computed from diagrams that have no internal $\phi$-lines.\footnote{
Diagrams involving the mean field (or one point function) $\langle\phi\rangle$, as shown in figure \ref{meanfieldfigure}, are of higher order in the tiny couplings $g, h_i$ and $Y$. This suppression is, however, compensated when the amplitude of the $\langle\phi\rangle$-oscillations is bigger than $\sim \alpha_i T^2/g$. We exclude this case, which may be realised during the early stage of reheating, as we are interested in the perturbative regime. 
}
Since all other fields are in equilibrium, $\Pi^-$ to this order is invariant under translations in time and depends only on the relative coordinate $x_1-x_2$. In principle it is straightforward to generalise the following considerations to a time dependent background \cite{Drewes:2012qw}, but the computation of $\Gamma$ in practice becomes much more difficult.
Thus, we can define the Fourier transforms
\begin{equation}
\Pi^\gtrless_\q(\omega)=\int d^4(x_1-x_2)\phantom{i}  e^{i\omega(t_1-t_2)} e^{-i\q(\textbf{x}_1-\textbf{x}_2)} \Pi^\gtrless(x_1-x_2)
.\end{equation}
Another quantity that we will need in the following is the spectral function $\Delta^-(x_1,x_2)$,
\begin{equation}
\Delta^-(x_1,x_2)=i\left(\langle\phi(x_1)\phi(x_2)\rangle-\langle\phi(x_2)\phi(x_1)\rangle\right)\label{deltaminusdefinition}
.\end{equation}
When $\Pi^-$ is translation invariant, the same applies to $\Delta^-$ \cite{Anisimov:2008dz}, and we can define the \textit{spectral density} $\rho_\q(\omega)$ as
\begin{equation}
\rho_\q(\omega)=-i\int d^4(x_1-x_2) \phantom{i} e^{i\omega(t_1-t_2)} e^{-i\q(\textbf{x}_1-\textbf{x}_2)} \Delta^-(x_1-x_2)\label{spectraldensitydefinition}
.\end{equation}
It can be expressed as \cite{Anisimov:2008dz}
\begin{eqnarray}\label{spectralfunction2}
\rho_{\q}(\omega)={-2{\rm Im}\Pi^R_{\q}(\omega)+2\omega\epsilon\over 
(\omega^2-m^2-\q^2-{\rm Re}\Pi^R_{\q}(\omega))^2+({\rm Im}\Pi^R_{\q}(\omega)+\omega\epsilon)^2}. 
\end{eqnarray}
The pole structure of $\rho_\q(\omega)$ in the complex $\omega$ plane determines the spectrum of propagating states in the plasma. Let $\hat{\Omega}_\q$ be a pole of $\rho_\q(\omega)$ with $\Omega_\q\equiv{\rm Re}\hat{\Omega}_\q$ and $\Gamma_\q\equiv2{\rm Im}\hat{\Omega}_\q$.
Both are temperature dependent because $\Pi^R_\q(\omega)$ depends on $T$.
In weakly coupled theories one usually observes the hierarchy
\begin{equation}\label{quasiparticle}
\Gamma_\q\ll\Omega_\q .
\end{equation}
In that case $\rho_\q(\omega)$ features peaks of width $\sim\Gamma_\q$ at energies $\omega\simeq\pm\Omega_\q$ which can be interpreted as {\it quasiparticle}-resonances.\footnote{Some authors use the expressions 
``quasiparticle approximation'' and ``narrow width approximation'' to
 refer to the limit $\Gamma_\q/\Omega_\q\rightarrow 0$, which corresponds to a spectral density (\ref{rhofree}).
We use these terms in a broad sense to refer to the situation of a small, but finite width described by (\ref{BWphi}), corresponding to unstable, but long-lived quasiparticles.} 
They may correspond to screened particles or collective excitations of the plasma. By ``screened particle'' we mean the pole that converges to $\hat{\Omega}_\q\rightarrow (\q^2+m^2)^{1/2}$ in the limit $T\rightarrow 0$.
It is clear from (\ref{spectralfunction2}) that, if (\ref{quasiparticle}) is fulfilled, 
the quasiparticle dispersion relation (or ``mass shell'') $\Omega_\q$ is essentially fixed by ${\rm Re}\Pi^R_{\q}(\omega)$ via the condition 
\begin{equation}\label{dispersionrelation}
\omega^2-\q^2-m^2-{\rm Re}\Pi^R_{\q}(\omega)=0, 
\end{equation}
while ${\rm Im}\Pi^R_{\q}(\omega)$ gives the thermal width.
The dispersion relation $\Omega_\q$ in general has a complicated $\q$-dependence. Only in special cases it can be parametrised by momentum independent ``thermal masses''\footnote{Such a parametrisation can be a good approximation if only a certain range of momentum modes is relevant. For very soft modes one may use the {\it plasma frequency}, i.e. the solution to (\ref{dispersionrelation}) for $\q=0$, for hard modes $\q\sim T$ one often uses the {\it asymptotic mass}., cf. (\ref{FermionPoles})}. 
${\rm Re}\Pi^R_\q(\omega)$ contains a zero temperature divergence that can be absorbed into the physical mass in the same way as in vacuum \cite{Boyanovsky:2004dj,Anisimov:2008dz}. In the following we interpret $m$ as physical mass, with ${\rm Re}\Pi^R_\q(\omega)$ finite. 
The physical properties of quasiparticles can be read off from a Breit-Wigner fit to (\ref{spectralfunction2}) near the pole, which is parametrised by $\Gamma_\q$ and $\Omega_\q$. 
\begin{equation}\label{BWphi}
\rho_{\textbf{q}}^{\rm BW}(\omega)=2\mathcal{Z}_{\q}\frac{\omega\Gamma_{\textbf{q}}}{(\omega^2-\Omega_{\textbf{q}}^2)^2+(\omega\Gamma_{\textbf{q}})^2} + \rho_{\textbf{q}}^{\rm cont}(\omega)
\end{equation}
$\rho_{\textbf{q}}^{\rm cont}(\omega)$ is the continuous part of $\rho_{\q}(\omega)$.
To obtain the correct residue, we introduced the parameter
\begin{equation}
\mathcal{Z}_\q=\left[1-\frac{1}{2\Omega_\q}\frac{\partial {\rm Re}\Pi^R_\q(\omega)}{\partial\omega}\right]^{-1}_{\omega=\Omega_\q}
.\end{equation}
Using the relation
\begin{equation}
\Pi^-_\q(\omega)=2i{\rm Im}\Pi^R_\q(\omega)\label{ImPiRvsPiMinus}
\end{equation}
we can express the thermal width as
\begin{equation}\label{GammaFormula}
\Gamma_\q=\mathcal{Z}_\q \frac{i}{2\Omega_\q}\Pi^-_\q(\Omega_\q).
\end{equation}
For $\phi$ we can set $\mathcal{Z}_\q=1$ because of the small inflaton coupling. For the other fields $\mathcal{X}_i$ it should be kept, cf. (\ref{OffShell}). 
$\Gamma_\q$ yields the total relaxation rate (or damping rate) for $\phi$ \cite{Anisimov:2008dz}, which plays the role of a production rate when the occupation of the mode $\q$ is below its equilibrium value and that of a dissipation rate in case it is above its equilibrium value, as in the case of reheating.  
$\Gamma_\q$ can be decomposed into gain- and loss rates $\Gamma_\q=\Gamma^>_\q-\Gamma^<_\q$, with $\Gamma_\q^\gtrless=i\Pi^\gtrless_\q(\Omega_\q)/(2\Omega_\q)$.
The relaxation for the coherent zero mode $\langle\phi\rangle$ is given by $\Gamma\equiv \Gamma_{\textbf{0}}$, see e.g. \cite{Morikawa:1986rp,Gleiser:1993ea,Ramsey:1997sa,Greiner:1998vd,Boyanovsky:2004dj,Yokoyama:2004pf,Anisimov:2008dz,Graham:2008vu,Gautier:2012vh}. 
We assume that due to the very weak inflaton coupling, thermal modifications of the $\phi$ dispersion relation are negligible.
Then there exists only one type of $\phi$ quasiparticle with $\Omega_\q\simeq\omega_\q=(\q^2+m^2)^{1/2}$. On the other hand, corrections to the thermal $\phi$-width $\Gamma_\q$ are important. 

In the models discussed in the following sections $\phi$ couples to other scalars $\chi_i$ and fermions $\Psi$ described by (\ref{L}).  
The spectral densities for $\chi_i$ and $\Psi_i$ are defined in analogy to (\ref{deltaminusdefinition}) and (\ref{spectraldensitydefinition}),
\begin{eqnarray}
\rho_{i \p}(p_0)&=&\int d^4(x_1-x_2)e^{i p_0(t_1-t_2)} e^{-i\p(\textbf{x}_1-\textbf{x}_2)}
\left(\langle\chi_i(x_1)\chi_i(x_2)\rangle-\langle\chi_i(x_2)\chi_i(x_1)\rangle\right)\\
(\uprho_{\p}(p_0))_{\alpha\beta}&=&\int d^4(x_1-x_2)e^{i p_0(t_1-t_2)} e^{-i\p(\textbf{x}_1-\textbf{x}_2)}
\left(\langle\Psi_{\alpha}(x_1)\bar{\Psi}_{\beta}(x_2)\rangle+\langle\bar{\Psi}_{\beta}(x_2)\Psi_{\alpha}(x_1)\rangle\right),\nonumber\\
\end{eqnarray}
where $_\alpha$ and $_\beta$ are spinor indices, which we suppress in the following. The different sign in the fermionic spectral density originates from the Grassmann nature of the fields. Self-energies $\Pi^-_i$ and $\Sigma^-$ for the $\chi_i$ and $\Psi$ can be defined in analogy to $\Pi^\pm$, but of course involve all interactions of the field under consideration.

In scalar theories at high temperature, there is usually only one type of quasiparticle that is relevant for transport \cite{LeB}. This justifies the (one) pole approximation
\begin{equation}\label{onepolerho}
\rho_{i \textbf{p}}^{\rm pole}(p_0)=\mathcal{Z}_{i \p}\frac{
i\mathcal{Z}_{i \p}\Pi_{i \p}^-(p_0)
}{(p_0^2-\Omega_{i \textbf{p}}^2)^2-(
\frac{\mathcal{Z}_{i \p}}{2p_0}\Pi^-_{i \p}(p_0)
)^2}. 
\end{equation}
We define the quantities $\mathcal{Z}_{i \p}$, $\Gamma_{i \textbf{p}}$ and $\Omega_{i \textbf{p}}$ for $\chi_i$ in analogy to $\mathcal{Z}_{\p}$, $\Gamma_{\textbf{p}}$ and $\Omega_{\textbf{p}}$ for $\phi$.
In the simplest case the $\chi_i$ dispersion relations can be parametrised as $\Omega_{i \textbf{p}}=(\textbf{p}^2+M_i^2(T))^{1/2}$, with momentum independent thermal masses $M_i(T)$.  
We will also use the Breit-Wigner approximation 
\begin{equation}\label{BW}
\rho_{i \textbf{p}}^{\rm BW}(p_0)=2\mathcal{Z}_{i \p}\frac{p_0\Gamma_{i \textbf{p}}}{(p_0^2-\Omega_{i \textbf{p}}^2)^2+(p_0\Gamma_{i\textbf{p}})^2} + \rho_{i \textbf{p}}^{\rm cont}(p_0)
\end{equation}
and its ``zero width limit'', 
\begin{equation}\label{rhofree}
\rho_{i \textbf{p}}^{\rm 0}(p_0)=2\pi\mathcal{Z}_{i \p}{\rm sign}(p_0)\delta(p_0^2-\Omega_{i \textbf{p}}^2) + \rho_{i \textbf{p}}^{\rm cont}(p_0)
.\end{equation}
Here $\Omega_{i \textbf{p}}={\rm Re}\hat{\Omega}_{i \textbf{p}}$ and $\Gamma_{i \textbf{p}}=2{\rm Im}\hat{\Omega}_{i \textbf{p}}$  are the quasiparticle energy and thermal width for $\chi_i$ excitations with spacial momentum $\textbf{p}$.
The continuous contribution due to multiparticle states is often suppressed in weakly coupled theories, see \cite{Boyanovsky:2004dj,Gautier:2012vh} for some discussion. Therefore it is often neglected.
The fermionic spectral density $\uprho_\p(p_0)$ is more complicated, we discuss it in section \ref{yukawasection}.

\begin{figure}
  \centering
    \includegraphics[width=16cm]{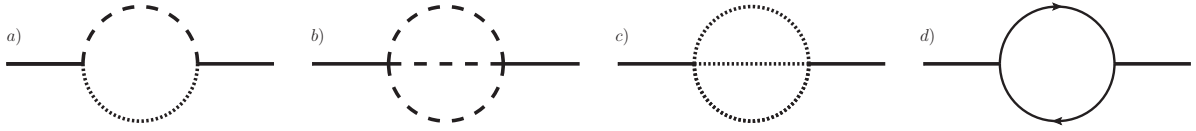}
    \caption{Relevant Feynman diagrams, solid lines represent $\phi$, dashed lines $\chi_1$, dotted lines $\chi_2$, solid lines with arrow $\Psi$.\label{diagrams}}
\end{figure}


\section{Example I: blocking by thermal masses}\label{example1}
We first discuss the dissipation into bosons.
We consider a situation where the interactions of $\phi$ can effectively be described by the following model
\begin{eqnarray}\label{L1}
\mathcal{L}&=& 
\frac{1}{2}\partial_{\mu}\phi\partial^{\mu}\phi
-\frac{1}{2}m^{2}\phi^{2}
+ \sum_{i=1}^2 \left(
\frac{1}{2}\partial_{\mu}\chi_{i}\partial^{\mu}\chi_{i}
-\frac{1}{2}m_{i}^{2}\chi_{i}^{2}-\frac{\lambda_i}{4!}\chi_i^4
\right)
-g\phi\chi_{1}\chi_{2}
+ \mathcal{L}_{\rm bath}.\nonumber\\
\end{eqnarray}
Here $\phi$ is the inflaton, $\chi_i$ are two other scalars with $m_i\ll m$ and 
$\mathcal{L}_{\rm bath}$ represents all other fields in the primordial plasma.
Using the real time formalism of thermal field theory (see e.g. \cite{LeB}), the contribution to $\Gamma$ from the diagram shown in figure \ref{diagrams}a) is obtained from\footnote{See e.g.  \cite{Drewes:2010pf}.} 
\begin{equation}\label{trilinear}
\Pi^-_\q(\omega)=-ig^2\int\frac{d^4p}{(2\pi)^4}\left(1+f_B(p_0)+f_B(\omega-p_0)\right)\rho_{1 \textbf{p}}(p_0)\rho_{2 \q-\textbf{p}}(\omega-p_0)
.\end{equation}
Here $f_B(\omega)=(e^{\omega/T}-1)^{-1}$ is the Bose-Einstein distribution.
We will use different approximations to the spectral densities $\rho_i$.

\begin{figure}
  \centering
    \includegraphics[width=9cm]{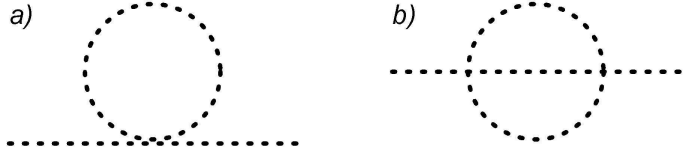}
    \caption{Diagrams contributing to the $\chi_i$ self-energy up to order $\lambda_i^2$.\label{phi4}}
\end{figure}
For the quartic $\chi_i$-self interaction in (\ref{L1}) the $\chi_i$-dispersion relations in the plasma can at leading order in $\lambda_i$ be parametrised as $\Omega_i=(\p^2+M_i^2)^{(1/2)}$, with momentum independent effective thermal masses that are generated by diagram shown in figure \ref{phi4}a)\footnote{The thermal mass correction here is $\sim\lambda^{1/2}T$ rather than $\sim\lambda T$ because it arises from a local (tadpole) diagram.}  
\begin{equation}\label{thermalmass}
M_{i}^{2}=m_{i}^{2}+\lambda_{i}\int\frac{d^{3}\textbf{p}}{(2\pi)^{3}}\frac{f_{B}(\bath{\omega}_{i})}{2\bath{\omega}_{i}}\approx m_{i}^{2}+\frac{\lambda_{i}}{24}T^{2}
.\end{equation}
Here we have suppressed the explicit temperature dependence of $M_i(T)$ \footnote{In the following we always use capital letters for effective masses and small letters for vacuum masses, often without making the dependence on $T$ explicit.}.
We assumed that the zero temperature part of the tadpole diagram has already been absorbed into $m_i$ by an appropriate renormalisation procedure\footnote{The dispersive part of the self-energy can only be absorbed into $m_i$ at a particular $T$ and appears explicitly for all other temperatures. It is common to impose renormalisation conditions at $T=0$. Then $m_i$ has the meaning of a physical mass in vacuum. The $T$-dependent part of the dispersive self-energy is finite and no counter terms in addition to the $T=0$ case are needed \cite{LeB}.}. The momentum independence of the diagram in in figure \ref{phi4}a) also implies $\mathcal{Z}_i=1$ at leading order in $\lambda_i$.

\subsection{Zero width approximation for quasiparticles}
\begin{figure}
  \centering
    \includegraphics[width=10cm]{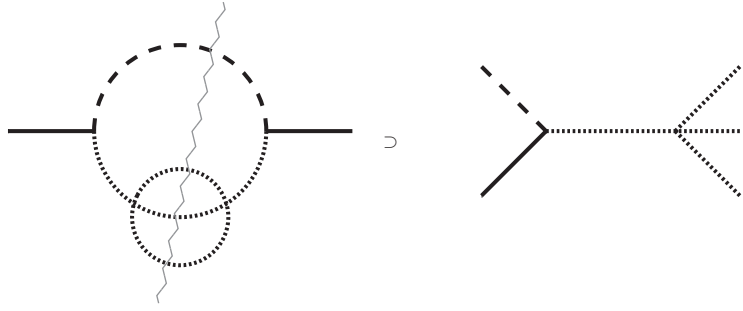}
    \caption{The self-energy $\Pi^-=2i{\rm Im}\Pi^R$ can be calculated from the sum of all cuts at a given order. 
At finite temperature, cut propagators can correspond to incoming or outgoing lines \cite{Bedaque:1996af}. For real scalar fields, a cut propagator that acts as incoming particle carries a factor $f_B$ (representing the abundance of scattering partners in the plasma), one that acts as outgoing particle has a factor $1+f_B$ (leading to Bose enhancement).  
The cut shown in this figure, for instance, includes $2\leftrightarrow 3$ scatterings.
Solid lines represent $\phi$, dashed lines $\chi_1$, dotted lines $\chi_2$.\label{cutting}}
\end{figure}

\begin{figure}
  \centering
    \includegraphics[width=16cm]{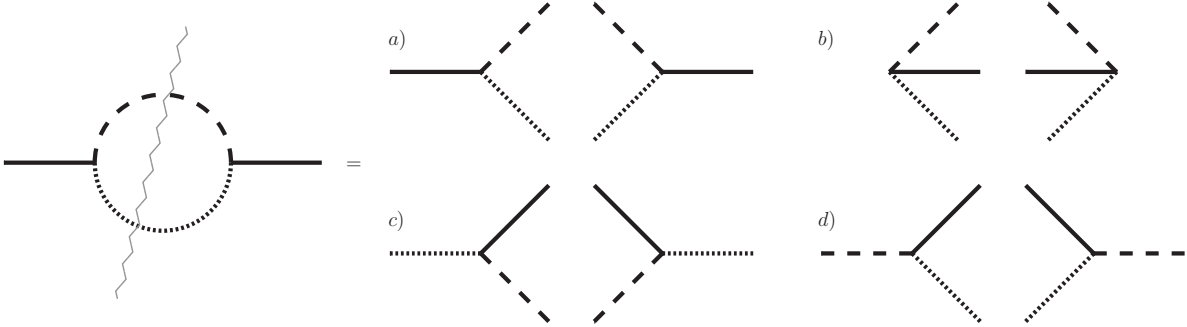}
    \caption{Different cuts through the leading order contribution to $\Pi^-$ from the trilinear interaction; solid lines represent $\phi$, dashed lines $\chi_1$, dotted lines $\chi_2$.\label{weldoncuts}}
\end{figure}

The structure of (\ref{trilinear}) is general for leading order diagrams. The product of spectral densities for the internal lines sorts out the kinematics of contributing processes, while the distribution functions account for the stochastic weight in the ensemble. 
When plugging in (\ref{rhofree}) for the spectral density, 
(\ref{trilinear}) is only non-zero when both of the particles in the loop can be on-shell simultaneously.
This is the case when the submanifolds defined by the $\delta$-functions in (\ref{rhofree}) intersect somewhere in the integration volume. 
It is in full analogy to the optical theorem and cutting rules at zero temperature \cite{Bedaque:1996af}. In vacuum, the total decay rate of a particle can be computed by cutting the self-energy in all possible ways and requiring that all cut lines must be on-shell. 
The different cuts are then interpreted as physical processes contributing to the decay rate, with the cut propagators as outgoing particles (decay products).
The difference at finite temperature is that the cut propagators can also act as incoming particles, cf. figures \ref{cutting} and \ref{weldoncuts}. The physical reason is that the medium provides scattering partners for $\phi$. 
In the Schwinger-Keldysh formalism, these processes are automatically taken into account in the calculation of  $\Pi^-$.  
Integration of (\ref{trilinear}) with (\ref{rhofree}) for $\q=0$ yields the known expression 
\begin{eqnarray}\label{DiplomFormula}
\lefteqn{\Pi^{-}_{\textbf{0}}(\omega)=-ig^{2}\int\frac{d^{3}\p}{(2\pi)^{2}}\frac{1}{4\Omega_{2}\Omega_{1}}}\nonumber\\
&\times&\bigg[\Big(\big(f_{1}+1\big)\big(f_{2}+1\big)-f_{1}f_{2}\Big)\Big(\delta(\omega-\Omega_1-\Omega_2)-\delta(\omega+\Omega_1+\Omega_2)\Big)\nonumber\\
&&+\phantom{X}\Big(\big(f_1+1\big)f_2-\big(f_2+1\big)f_1\Big)\Big(\delta(\omega-\Omega_1+\Omega_2)-\delta(\omega+\Omega_1-\Omega_2)\Big)\bigg]
,\end{eqnarray}
with $\Omega_i=\Omega_{i \p}$ and $f_i=f_B(\Omega_i)$.
We have neglected the continuum part of (\ref{rhofree}) in (\ref{DiplomFormula}), which is self-consistent because its contribution is of the same order as the deviation of $\mathcal{Z}_i$ from 1.
The different $\delta$-functions in (\ref{DiplomFormula}) arise from different cuts through the self-energy shown in figure \ref{diagrams}a) \cite{Weldon:1983jn}. They are displayed in figure \ref{weldoncuts}. The term proportional to $\delta(\omega-\Omega_1-\Omega_2)$ in the first line describes decays and inverse decays $\phi\leftrightarrow\chi_1\chi_2$, see figure \ref{weldoncuts}a). The two terms in the prefactor can be identified with the thermodynamic probabilities $(f_{1}+1)(f_{2}+1)$ for a decay (the $f_i$ are due to Bose enhancement as the $\chi_i$ are bosons) and $f_{1}f_{2}$ for an inverse decay; they obey the detailed balance relation. The $\delta(\omega+\Omega_1+\Omega_2)$ term comes with the same prefactor, but only contributes for $\omega<0$; it represents the generation of $\phi\chi_1\chi_2$ from the vacuum and the inverse process, which give no on-shell contribution.
The terms in the second line can be interpreted correspondingly: they represent processes $\chi_2\leftrightarrow\phi\chi_1$ and $\chi_1\leftrightarrow\phi\chi_2$, see figure \ref{weldoncuts}c) and d).

Then the integral (\ref{DiplomFormula}) can be evaluated analytically \cite{Boyanovsky:2004dj,Anisimov:2008dz} as
\begin{eqnarray}\label{dampingformel}
\Pi^{-}_{\textbf{0}}(\omega)=2i\big(\mathcal{D}^{[g]}_{\textbf{0}}(\omega)+\mathcal{S}^{[g]}_{\textbf{0}}(\omega)-\mathcal{D}^{[g]}_{\textbf{0}}(-\omega)-\mathcal{S}^{[g]}_{\textbf{0}}(-\omega)\big)
\end{eqnarray}
with
\begin{eqnarray}
\mathcal{D}^{[g]}_{\textbf{0}}(\omega)&=&-\frac{g^2}{8\pi}
\frac{\sqrt{w_1^2-M_1^2}}{w_2+w_1}
\big(1+f_B(w_1)+f_B(w_2)\big)
\theta(\omega-M_1-M_2)\theta(\omega)
\end{eqnarray}
\begin{eqnarray}
\mathcal{S}^{[g]}_{\textbf{0}}(\omega)&=&-\frac{g^2}{8\pi}
\frac{\sqrt{w_1^2-M_1^2}}{w_2-w_1}
\big(f_B(w_2)-f_B(w_1)\big)\theta(-\omega-M_1+M_2)\theta(\omega)
,\end{eqnarray}
where we have used the abbreviations
\begin{equation}
w_1=\frac{\omega^2+M_1^2-M_2^2}{2\omega} \ \ , \ \ w_2=\frac{\omega^2-M_1^2+M_2^2}{2\omega}.
\end{equation}
The terms have clear physical interpretations. 
$\mathcal{D}^{[g]}_\q(\omega)$ is the total contribution from the decays and inverse decays $\phi\leftrightarrow\chi_{1}\chi_{2}$. 
The $\theta$-function shows that such processes are only allowed if $M\geqslant M_1+M_2$, as suggested by on-shell quasiparticle kinematics.  
$\mathcal{S}_\q(\omega)$ is the contribution from processes $\chi_i\phi\leftrightarrow\chi_j$, which give rise to Landau damping.  It comes with a $\theta$-function that requires $|M_1-M_2|\geqslant M$.
In between there is a kinematic region in which neither of these processes is kinematically allowed.

Let us summarise this behaviour on more general grounds.
Let us assume that there is an interaction that gives thermal masses $M_i\sim\alpha_i T$ to the decay products, where $\alpha_i$ is a dimensionless small parameter that we use symbolically to indicate scales in a weakly coupled system. For a particular interaction it has to be related to the coupling constant.
For instance, to parametrise the thermal mass (\ref{thermalmass}) as $\alpha_i T$, we have to identify $\alpha_i\equiv(\lambda_i/24)^{1/2}$, see  (\ref{thermalmass}).\footnote{The unusual square-root scaling is due to the tadpole diagram in figure \ref{phi4}. For fermions with hard momenta one has to identify $\alpha_i\equiv\upalpha/2$, see (\ref{FermionPoles}).}  
Since thermal corrections to the inflaton dispersion relation are negligible and its width is small, we can set $M\simeq m$.
The $\chi_i$ vacuum masses $m_i$ are much smaller than $M\simeq m$. However, due to $\alpha_i\gg g/m$, they grow much faster with $T$ and we can define critical temperatures 
\begin{eqnarray}
\begin{tabular}{c c c}
$m^2=(M_1(T_c)+M_2(T_c))^2$ &, & $m^2=(M_1(\tilde{T}_c)-M_2(\tilde{T}_c))^2$  
\end{tabular}
.\end{eqnarray}
We can estimate the critical temperatures as $T_c\sim \frac{m}{\alpha_1+\alpha_2}$ and $\tilde{T}_c\sim \frac{m}{|\alpha_1-\alpha_2|}$.
For $T<T_c$, $\phi$ can reheat the plasma by decay, for $T>\tilde{T}_c$ by Landau damping.
If it can be justified to use (\ref{dampingformel}) to compute the total relaxation rate $\Gamma$, $T_c$ indeed poses an upper bound on the temperature because the primordial plasma can never reach temperatures $T>\tilde{T}_c$, where Landau damping would be efficient. 
However, with the $\chi_i$ self-interactions (\ref{L1}), the trilinear coupling $g\phi\chi_1\chi_2$ allows for scatterings $\phi\chi_i\leftrightarrow\chi_j\chi_j\chi_j$, see figure \ref{cutting}.
These require the exchange of an intermediate $\chi_j$ quantum\footnote{The $h_i\phi\chi_i^3/4!$ interaction in (\ref{L}) also allows for scatterings without intermediate state, which we will discuss in the following section \ref{QuarticInteraction}.}. In the region $T_c<T<\tilde{T}_c$, this intermediate state must be off-shell. In addition, there is a suppression by the coupling constant. In spite of that they can be relevant at high temperature because both of these suppressions may be compensated by the large occupation numbers in the plasma. Whether or not the approximation (\ref{dampingformel}) is justified has to be decided on a case by case basis. 

\subsection{Inclusion of thermal widths and off-shell transport}\label{OnShellvsOffShell}
In the Schwinger-Keldysh formalism, all contributions are taken into account consistently when  the $\phi$-self-energy $\Pi^-_\q(\omega)$ is calculated at higher order in the couplings $\alpha_i$. 
A consistent computation of $\Pi^-_\q(\omega)$ to a given order in all couplings automatically includes all processes that contribute to $\Gamma$ at that order. 
This may require resummation \cite{Braaten:1989mz} to fix the so-called \textit{breakdown of perturbation theory}\footnote{This terminology is somewhat misleading; what breaks down here is only the loop expansion, which is not identical to the perturbative expansion and not controlled by a small parameter. Perturbation theory still works if all quantities are consistently calculated at a given order in a small parameter, which may require resummation \cite{LeB}. The fact that such calculations are often referred to as ``non-perturbative'' is similarly misleading, as they are still based on a (resummed) perturbative expansion in a small parameter. This is e.g. in contrast to lattice computations in strongly coupled systems, where no small expansion parameter exists.} at high temperature \cite{LeB}. 
The need for resummation is usually related to infrared or collinear divergences.
Infrared problems at finite temperature mathematically are related to the divergent behaviour of the Bose-Einstein distribution function $f_B$ for small arguments. Physically, they can be understood easily: At high temperatures, the density of potential scattering partners in the plasma becomes so large that the probability for multiple scatterings within some time unit is of the same order of magnitude as that for a single interaction.
Collinear enhancement in thermal field theory \cite{Anisimov:2010gy,Garbrecht:2013gd} can occur when the decaying particle is relativistic and the decay products move parallel for a long time in the rest frame of the bath, giving them enough time for multiple exchange of force carriers. 

\begin{figure}
  \centering
    \includegraphics[width=6cm]{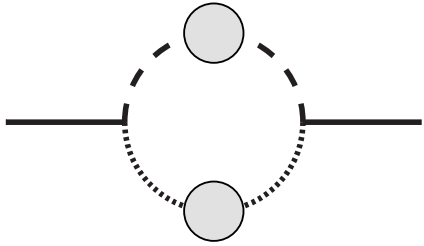}
    \caption{Diagrammatic representation of (\ref{OffShell}), i.e. (\ref{trilinear}) evaluated with the full (resummed) spectral density; solid lines represent $\phi$, dashed lines $\chi_1$, dotted lines $\chi_2$, the gray blobs symbolise self-energy insertions.\label{dressed}}
\end{figure}
\begin{figure}
  \centering
    \includegraphics[width=12cm]{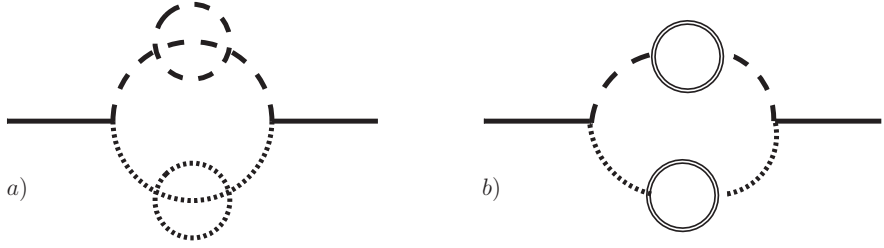}
    \caption{Finite $\Gamma_i$ contributions to (\ref{OffShell}) for $\chi_i$ interactions as in (\ref{L1}) in diagram $a)$ and for $\chi_i$ interactions as in (\ref{L2}) in diagram $b)$. Solid lines represent $\phi$, dashed lines $\chi_1$, dotted lines $\chi_2$,  double lines $\xi$.\label{fiveloop}}
\end{figure}

For the Lagrangian (\ref{L1}) the full relaxation rate $\Gamma$ can be calculated from (\ref{trilinear}) by using dressed spectral densities instead of (\ref{rhofree}).
This is always justified when diagrams as shown in figure \ref{dressed} dominate over vertex corrections and ``ladder diagrams'' that contribute to $\Pi^-_\q(\omega)$, which are not included in (\ref{trilinear}). 
For simplicity, we have constructed the Lagrangian (\ref{L1}) to suppress corrections to the $g$ vertex and ladder diagrams\footnote{Ladder diagrams have e.g. been studied in \cite{BasteroGil:2010pb} for scalars and in \cite{Anisimov:2010gy,Garbrecht:2013gd} for fermions.}. In this case the effect of scattering processes encoded in the diagrams shown in figure \ref{fiveloop} can be parametrised by the thermal widths $\Gamma_i$ of the $\chi_i$.

Using (\ref{GammaFormula}) to replace $\frac{\mathcal{Z}_{i \p}}{2p_0}\Pi^-_{i \p}(p_0)$ by $\Gamma_i$ in the denominator of (\ref{onepolerho}), we can perform the $p_0$ integration in (\ref{trilinear}) with Cauchy's theorem. 
We need to keep the full $\omega$ dependence of the remaining $\Pi_j^-$ in the numerator to account for $\rho_{i \textbf{p}}^{\rm cont}(p_0)$, especially near the regions where the arguments of the Bose-Einstein distributions vanish. This corresponds to the approximation (\ref{onepolerho}).
Then $\Pi^-_\q(\omega)$ for $\q=0$ reads 
\begin{eqnarray}
\Pi^{-}_{\textbf{0}}(\omega)&=&\frac{g^{2}}{2}\int\frac{d|\textbf{p}|}{(2\pi)^{2}}\textbf{p}^{2}\mathcal{Z}_{1}\mathcal{Z}_{2}\nonumber\\
&\times&\Bigg[
\frac{\left(\mathcal{Z}_2\Pi^{-}_{2}(\omega-\Omega_{1})-\epsilon2i(\omega-\Omega_{1})\right)\left(1+f_{B}(\Omega_{1})+f_{B}(\omega-\Omega_{1})\right)}{\hat{\Omega}_{1}\left(\left((\omega-\hat{\Omega}_{1})^{2}-\Omega_{2}^{2}\right)^{2}+\left(\frac{i}{2}\mathcal{Z}_2\Pi^{-}_{2}(\omega-\hat{\Omega}_{1})+\epsilon(\omega-\Omega_{1})\right)^{2}\right)}\nonumber\\
&&+\frac{\left(\mathcal{Z}_2\Pi^{-}_{2}(\omega+\Omega_{1})-\epsilon2i(\omega+\Omega_{1})  \right)\left(
f_{B}(\Omega_{1})
-f_{B}(\omega+\Omega_{1})
\right)}{\hat{\Omega}_{1}^{*}\left(\left((\omega+\hat{\Omega}_{1}^{*})^{2}-\Omega_{2}^{2}\right)^{2}+\left(\frac{i}{2}\mathcal{Z}_2\Pi^{-}_{2}(\omega+\hat{\Omega}_{1}^{*})+\epsilon(\omega+\Omega_{1}) \right)^{2}\right)}\nonumber\\
&&+\frac{\left(\mathcal{Z}_1\Pi^{-}_{1}(\omega+\Omega_{2})-\epsilon2i(\omega+\Omega_{2}) \right)\left(f_{B}(\Omega_{2})
-
f_{B}(\omega+\Omega_{2})
\right)}{\hat{\Omega}_{2}\left(\left((\omega+\hat{\Omega}_{2})^{2}-\Omega_{1}^{2}\right)^{2}+\left(\frac{i}{2}\mathcal{Z}_1\Pi^{-}_{1}(\omega+\hat{\Omega}_{2})+\epsilon(\omega+\Omega_{2})\right)^{2}\right)}\nonumber\\
&&+\frac{\left(\mathcal{Z}_1\Pi^{-}_{1}(\omega-\Omega_{2})-\epsilon2i(\omega-\Omega_{2}) \right) \left(1+f_{B}(\Omega_{2})+f_{B}(\omega-\Omega_{2})\right)}{\hat{\Omega}_{2}^{*}\left(\left((\omega-\hat{\Omega}_{2}^{*})^{2}-\Omega_{1}^{2}\right)^{2}+\left(\frac{i}{2}\mathcal{Z}_1\Pi^{-}_{1}(\omega-\hat{\Omega}_{2}^{*})+\epsilon(\omega-\Omega_{2})\right)^{2}\right)}\Bigg].\label{OffShell}
\end{eqnarray}
Here we have again suppressed the spacial momentum index. 
 The $\chi_i$ self-energies $\Pi^-_i$ are all evaluated for spacial momenta $\textbf{p}$.  
The infinitesimal parameter $\epsilon$ has to be kept to regularise the integrand in case a pole lies outside the support of the $\Pi_i^-$ in the numerator\footnote{The support of the full spectral self-energy covers the entire energy axis at $T\neq0$ because it contains cuts corresponding to scattering processes without kinematic  branchcut. 
However, at a given order in perturbation theory, the support may be limited due to such branchcuts, cf. (\ref{dampingformel}), (\ref{xiIntSelfEn}) and (\ref{A}).}. 
We kept the factors $\mathcal{Z}_i$, which are in good approximation unity for the quartic self-interaction in (\ref{L1}), because we will need them in the following section.
The result holds to first order in $\Gamma_i/\Omega_j$. 
In the limit $\Gamma_i/\Omega_i\rightarrow 0$ it reproduces the analytic formula (\ref{dampingformel}). In contrast to  (\ref{dampingformel}), it does not vanish for $T_c<T<\tilde{T}_c$ because it includes processes in which the $\chi_i$ are off-shell. 
Physically this means that there exists no strictly stable (quasi)particles at finite temperature because even those particles that would be stable in vacuum can be annihilated in a scattering with quanta from the bath.
\footnote{This behaviour has been found in the thermal production rate of right handed neutrinos at high temperature \cite{Anisimov:2010gy}, which is relevant for a consistent treatment of leptogenesis, and chemical equilibration in hot plasmas \cite{Garbrecht:2008cb}. A similar effect has also been described in de Sitter spacetime even in the absence of a thermal plasma \cite{Boyanovsky:2004ph}.}

\subsection{The kinematic regimes}\label{numerical}
\begin{figure}
  \centering
    \includegraphics[width=12cm]{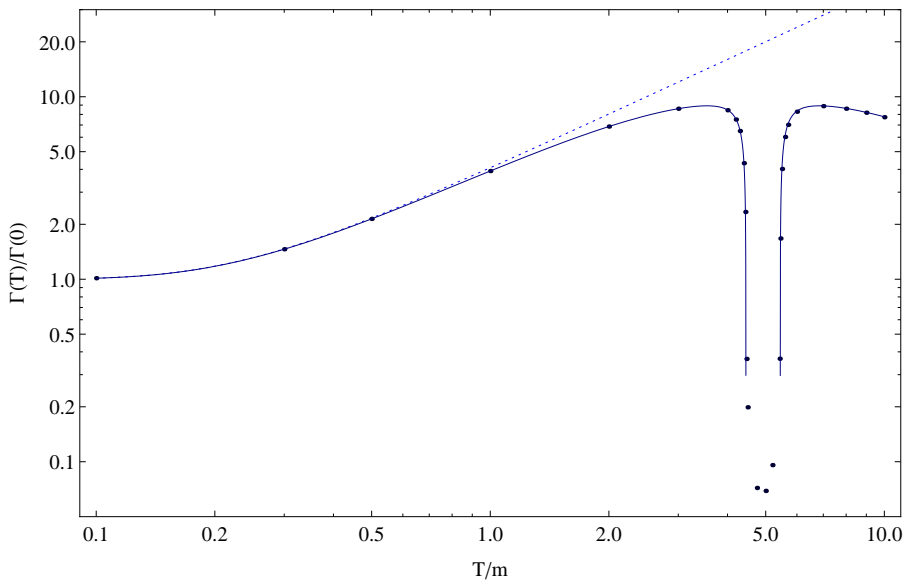}
    \caption{The rate $\Gamma$ in the model defined by (\ref{L1}) as a function of $T$, normalised to its zero temperature value.
We set $\lambda_1=0.01$, $\lambda_2=1$, $m_1=m_2=0.001 m$. 
The solid line is calculated from (\ref{dampingformel}), it takes into account thermal mass corrections for $\chi_i$, but neglects the thermal widths.
The dashed light blue line is the result one obtains when also neglecting thermal masses. 
The dark blue dots are numerical evaluations of (\ref{OffShell2}), which also take into account thermal $\chi_i$-widths.  
\label{offshellplot}}
\end{figure}

We consider the case $\lambda_2\gg\lambda_1$ to study the temperature regime $T>T_c$.
This e.g. qualitatively resembles a situation in which one of the fields is charged under some gauge group while the other one is not. In the SM a similar situation is realised for the Higgs coupling to electrons and neutrinos.
It also resembles a coupling to active and sterile neutrinos.
Then off-shell transport due to the tiny $\chi_1$-width is negligible (while we keep the $\chi_2$ width) and we can use the functions $\mathcal{D}^{[g]}_\q$, $\mathcal{S}^{[g]}_\q$ introduced in  (\ref{dampingformel}) to approximate (\ref{OffShell}) by
\begin{eqnarray}
\Pi^{-}_{\textbf{0}}(M)&=&\frac{g^{2}}{2}\int\frac{d|\textbf{p}|}{(2\pi)^{2}}\frac{\textbf{p}^{2}}{\Omega_1}\Bigg(
\frac{\Pi^{-}_{2}(M-\Omega_{1})\left(1+f_{B}(\Omega_{1})+f_{B}(M-\Omega_{1})\right)}{\left((M-\Omega_{1})^{2}-\Omega_{2}^{2}\right)^{2}+\left(
\Gamma_2\Omega_2
\right)^{2}}\nonumber\\
&&-\frac{\Pi^{-}_{2}(M+\Omega_{1})\left(f_{B}(M+\Omega_{1})-f_{B}(\Omega_{1})\right)}{\left((M+\Omega_{1})^{2}-\Omega_{2}^{2}\right)^{2}+\left(
\Gamma_2\Omega_2
\right)^{2}}\Bigg)\nonumber\\
&&+i\big(\mathcal{D}^{[g]}_{\textbf{0}}(M)+\mathcal{S}^{[g]}_{\textbf{0}}(M)-\mathcal{D}^{[g]}_{\textbf{0}}(-M)-\mathcal{S}^{[g]}_{\textbf{0}}(-M)\big)\label{OffShell2}
,\end{eqnarray}
The integral in (\ref{OffShell2}) has to be evaluated numerically. 
This requires knowledge of the self-energy $\Pi^-_2$ for arbitrary momenta and off-shell energies.
We obtain these by creating a numerical table, based on the equations given in appendix \ref{appendix}, which we then insert into (\ref{OffShell2}).
The result for $\Gamma$ obtained from (\ref{OffShell2}) is shown and compared to the on-shell result from (\ref{dampingformel}) in figure \ref{offshellplot}.

The behaviour of (\ref{dampingformel}) can be understood easily. 
For $T<T_c$ $\phi$ dissipates its energy via decays and inverse decays.
The temperature $T_{max}$ at which this is most efficient is determined by the competition between amplification by Bose enhancement and the decreasing phase space at increasing temperature.
For $T>T_c$ the decay is kinematically forbidden and (\ref{dampingformel}) is zero.
For negligible vacuum masses, $T_c$ can be estimated as $T_c\simeq m\sqrt{24}/(\sqrt{\lambda_1}+\sqrt{\lambda_2})$. 
At $\tilde{T}_c\simeq m\sqrt{24}/|\sqrt{\lambda_1}-\sqrt{\lambda_2}|$ the processes $\phi\chi_1\leftrightarrow\chi_2$ become kinematically allowed and give rise to Landau damping.

The full relaxation rate, calculated from (\ref{OffShell2}), shows a very similar behaviour. It also increases linearly with $T$ due to the effect of Bose enhancement between $T$ and a temperature $T_{max}' \sim T_{max}$, where (\ref{OffShell2}) has a local maximum.
Beyond $T_{max}'$ the relaxation rate again decreases with $T$, as the shrinking phase space overpowers the effect of Bose enhancement.
In the region $T_c<T<\tilde{T}_c$, where at least one of the intermediate states in any process contributing to the relaxation, we observe $\Gamma\neq0$ due to off-shell transport. 
$\Gamma$ receives contributions from scatterings encoded in the loop diagram shown in figure \ref{cutting}. 
As an example we show one scattering process ($\phi\chi_1\leftrightarrow\chi_2\chi_2\chi_2$) obtained from the cut in figure \ref{cutting}, but the loop also contains scatterings  
$\phi\chi_2\leftrightarrow\chi_1\chi_2\chi_2$,  $\phi\chi_2\chi_2\leftrightarrow\chi_1\chi_2$ and $\phi\chi_1\chi_2\leftrightarrow\chi_2\chi_2$ as well as (if kinematically allowed) decays and inverse decays such as $\phi\leftrightarrow \chi_1\chi_2\chi_2\chi_2$ or $\chi_1\leftrightarrow \phi\chi_2\chi_2\chi_2$.
These are suppressed by an additional coupling $\lambda_2$, the virtuality of the intermediate $\chi_2$ quasiparticle and the required phase space. They are not included in (\ref{dampingformel}). 
On the other hand, they are amplified by the infrared behaviour of the $f_B$ due to Bose enhancement.
The resulting overall suppression is very efficient not only compared to $\Gamma(T_{max})$, but even relative to the decay width in vacuum. Thus, if this scenario is realised in nature, perturbative reheating indeed effectively stopped at $T_c$, as claimed in \cite{Kolb:2003ke} (the plasma would, however, be heated much faster than calculated in that article as the authors neglected the enhancement by Bose enhancement for $T<T_c$). 

The effective suppression is a consequence of the interaction Lagrangian (\ref{L1}).
The suppression in the regime $T_c<T<\tilde{T}_c$ is particularly strong for the interaction Lagrangian (\ref{L1}) for three reasons. 
First, it does not contain any interaction that couples $\chi_1$ and $\chi_2$ to each other.\footnote{We indirectly assumed that they communicate in some way via $\mathcal{L}_{\rm bath}$  by assigning the same temperature $T$ to both. The results in this section remain valid as long as the kinetic equilibration between $\chi_1$ and $\chi_2$ happens faster than $1/\Gamma$.} If $\chi_1$ and $\chi_2$ were e.g. charged under the same gauge group, $\Gamma$ would receive contributions from ``ladder diagrams'' similar to the one for fermions shown in figure \ref{ladders}. 
We avoided such terms to obtain the transparent expression (\ref{OffShell}), which is analytic up to a one-dimensional integral. 
It is known from other examples that such contributions can qualitatively change the behaviour. For instance, the production rate of right handed neutrinos from a thermal plasma at tree level exhibits a kinematically forbidden temperature region similar to the interval $T_c<T<\tilde{T}_c$ discussed here \cite{Giudice:2003jh}.
There ladder diagrams do contribute, and in \cite{Anisimov:2010gy} it has been found that their resummation can almost entirely overcome the suppression in the ``forbidden region''. 
For a choice of $\mathcal{L}_{\mathcal{X}}$ that yields contributions to $\Gamma$ from ladder diagrams during reheating we expect a similar effect, though it is probably less pronounced than in the neutrino production rate due to the absence of the collinear kinematics that is crucial to the result found in \cite{Anisimov:2010gy} (cf. caption of figure \ref{ladders}).
Second, the quartic self interaction in (\ref{L1}) leads to a contribution to the dispersive part of the $\chi_i$-self-energy at order $\lambda_i$, cf. figure \ref{phi4}a), giving rise to a thermal mass $\propto\sqrt{\lambda_i}T$. In contrast to that, the lowest order contribution to the dissipative self-energy $\Pi_i^-$ is of order $\lambda_i^2$, cf. figure \ref{phi4}b). 
This disfavours off-shell transport if $\lambda_i$ is a small parameter: 
$\Pi^-_i$ appears in the numerator of (\ref{OffShell}) and scales as $\propto \lambda_i^2$.
The critical temperature $T_c$, on the other hand, depends on the coupling as $\lambda_i^{-1/2}$ because $M_i\propto \lambda_i^{1/2}$. The factors $f_B$ in the numerator of (\ref{OffShell}) enhance $\Gamma$ in the ``forbidden region'' more efficiently for a large $T_c$, thus small $\lambda_i$. This effect competes with the $\lambda_i^2$ suppression from $\Pi_i^-$ in the numerator.
For a gauge or Yukawa coupling the dispersive and dissipative part of the self-energy appear at the same order, and the thermal mass is proportional to the coupling (rather than its square root).
This allows the effect of Bose enhancement to be more efficient at $T_c$. 
Furthermore, the fact that the lowest order contribution to $\Pi_i^-$ comes from a two loop diagram leads to an additional suppression, which e.g. can be noticed in the small numerical prefactor in (\ref{SimpleScatterings}). This furthermore suppresses $\Gamma$ in the forbidden region via the $\Pi^-$ in the denominator of (\ref{OffShell})\footnote{A small $\Pi_i^-$ does not suppress $\Gamma$ in the regions where on-shell transport is kinematically allowed because in this case the integration volume includes narrow pole regions where the denominator of the integrand in (\ref{OffShell}) is of order $1/(\Pi_i^-)^2$ and the integral is of order one.}. 
Finally, the momentum independence of the thermal mass (\ref{thermalmass}) implies that even soft modes become very thermally heavy.
For these reasons the upper bound on the reheating temperature found in this section is a rather special case.

\section{Example II: no blocking by thermal masses}\label{example2}
We now consider a situation in which the coupling between $\phi$ and $\chi_i$ is the same as in (\ref{L1}), but the $\chi_i$ have different interactions. 
This illustrates how strongly the behaviour of $\Gamma$ depends on the interactions within the primordial plasma even if the inflaton coupling is the same.
We use the following model
\begin{eqnarray}\label{L2}
\mathcal{L}&=& 
\frac{1}{2}\partial_{\mu}\phi\partial^{\mu}\phi
-\frac{1}{2}m^{2}\phi^{2}-g\phi\chi_{1}\chi_{2}
+\frac{1}{2}\partial_{\mu}\xi\partial^{\mu}\xi-\frac{1}{2}m_\xi^{2}\xi^{2}
\\
&+& \sum_{i=1}^2 \left(
\frac{1}{2}\partial_{\mu}\chi_{i}\partial^{\mu}\chi_{i}
-\frac{1}{2}m_{i}^{2}\chi_{i}^{2}
-\mathfrak{g}_i\chi_i \xi^2
\right)
+ \mathcal{L}_{\rm bath}.\nonumber
\end{eqnarray}
Here $\xi$ is another real scalar field.
\begin{figure}
  \centering
    \includegraphics[width=12cm]{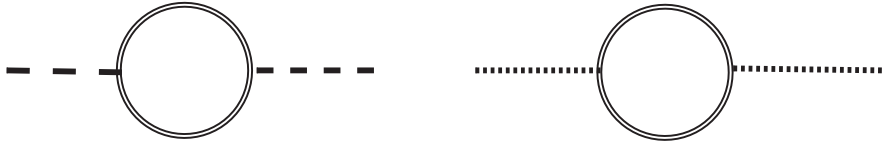}
    \caption{Leading order contributions to the $\chi_i$-self-energies in the model (\ref{L2}); we use the same notation as in figure \ref{fiveloop}.\label{xicontrib}}
\end{figure}
We can again use (\ref{trilinear}) to calculate $\Gamma$, but
in this case the dissipative and dispersive parts of the $\chi_i$ self-energies are both calculated from the one loop diagrams shown in figure \ref{xicontrib}.
The dissipative part can be found analytically and reads
\cite{Drewes:2013bfa} 
\begin{eqnarray}\label{xiIntSelfEn}
\lefteqn{ 
\Pi^-_{i,\mathbf{p}}=\frac{\mathfrak{g}_i^2}{16 i \pi |\mathbf{p}|} 
\Bigg[- \theta(-p^2) \Bigg( p_0 + T \log\bigg[
\frac{
f_B(p_0-\omega^+)f_B(-\omega^-)
}{
f_B(-\omega^+)f_B(p_0-\omega^-)
}
\bigg]\Bigg)
}
\nonumber\\
&+&
\theta\big(p^2 - (2m_\xi)^2\big) 
T \log\bigg[\frac{f_B(p_0-\omega^+)f_B(-\omega^-)}{f_B(-\omega^+)f_B(p_0-\omega^-)}\bigg]
 \Bigg]\nonumber
\end{eqnarray}
with the Bose-Einstein distribution $f_B(p_0)\equiv(e^{p_0/T}-1)^{-1}$ and
\begin{eqnarray}
\omega^\pm=\frac{p_0}{2}\pm{\rm sign}(p^2)\frac{|\textbf{p}|}{2}\sqrt{
1-\frac{(2m_\xi)^2}{p^2}}\nonumber
\end{eqnarray}
Since the dissipative and dispersive self-energies can be identified with real and imaginary part of the function $\Pi^R_i$, one can use (\ref{dispersionrelation}) and the Kramers-Kronig relations to find the following equation for $\Omega_i$\footnote{When evaluating (\ref{disprelfinder}) we drop the $T=0$ part of (\ref{xiIntSelfEn}), which would give rise to an infinite contribution that has to be removed by mass renormalisation. As for (\ref{thermalmass}), we assume that this term has already been absorbed into $m_i$.}
\begin{equation}
\Omega_i^2-\textbf{p}^2-m_i^2-\mathcal{P}\int\frac{d \omega}{2\pi i}\frac{\Pi_i^-(\omega)}{\omega-\Omega_i}=0.\label{disprelfinder}
\end{equation}

\begin{figure}
  \centering
    \includegraphics[width=12cm]{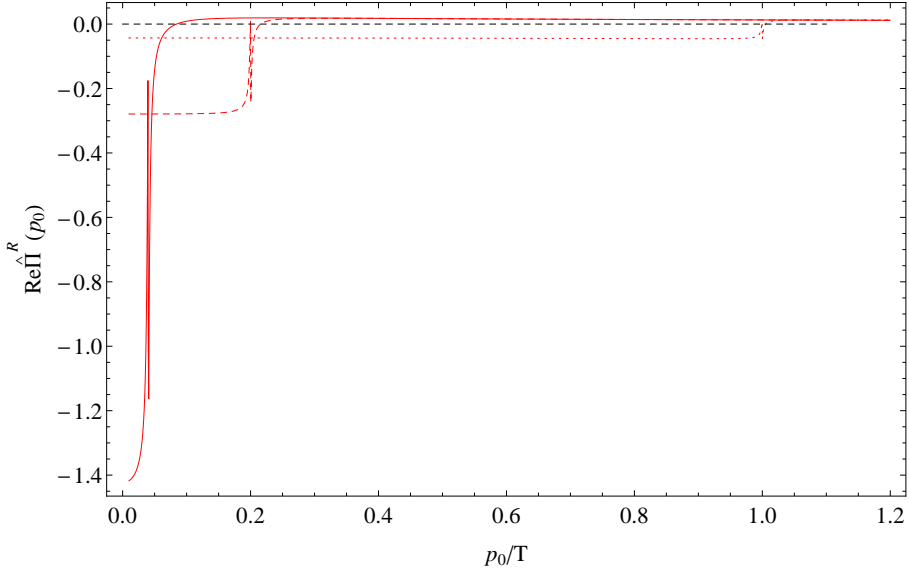}
    \caption{The dispersive part of the $\chi_i$-self-energy ${\rm Re}\hat{\Pi}^R_i(p_0)={\rm Re}\Pi^R_i(p_0)/\mathfrak{g}_i^2$ for $m_\xi/m_i=10^{-2}$ at $T=5m_i$ for $|\p|=m_i/5$ (solid line) and $|\p|=m_i$ (dashed line) and $|\p|=T=5m_i$ (dotted line) as a function of energy. For hard modes with $|\p|\gtrsim T$ and small coupling $\mathfrak{g}_i\ll1$ the solution of (\ref{disprelfinder}) is dominated by the $\p^2$. Then $\Omega_i$ lies in the regime where ${\rm Re}\Pi^R_i(p_0)$ is small and positive. However, for $m_i, |\p| \ll T$ the number of solutions for (\ref{disprelfinder}) and the sign of the thermal mass correction can vary. 
\label{RePiPlot}}
\end{figure}

The dispersive part is plotted in figure \ref{RePiPlot}. The most prominent feature of the function ${\rm Re}\Pi^R_\p(p_0)$ is the spike, which appears around $p_0^2=\p^2+(2m_\xi)^2$ because the pole in the last term in (\ref{disprelfinder}) passes the threshold of (\ref{xiIntSelfEn}).
The momentum dependence implies that the solution to (\ref{disprelfinder}) in this regime can in general not be parametrised by a momentum independent thermal mass.
It is sometimes argued that one can simply use an effective thermal mass that is determined by fixing the momentum to $T$, as most particles in a thermal plasma at $T\gg m_i$ have momenta $\sim T$. This argument fails in our case because the integral in (\ref{OffShell}) is not always dominated by the region around $|\textbf{p}|\sim T$ that maximises $\textbf{p}^2f_B(|\textbf{p}|)$, but by the regions around the poles.
The momentum scale in $\phi\rightarrow\chi_i\chi_j$ decays is set by the $\phi$ mass, the maximal momentum the decay products can have is $|\textbf{p}|=m/2$. For $T\gg m$ this corresponds to ``soft momenta''. 
That means that, for sufficiently high $T$, $\Gamma$ is sensitive to the shape of the $\chi_i$ spectral densities in the infrared regime, where the principal value in (\ref{disprelfinder}) dominates over $\p^2+m_i^2$. 
In this regime the  pole structure can be rather complicated. Physically this happens because particles with soft momenta can be strongly affected by the medium.
 
\begin{figure}
  \centering
    \includegraphics[width=12cm]{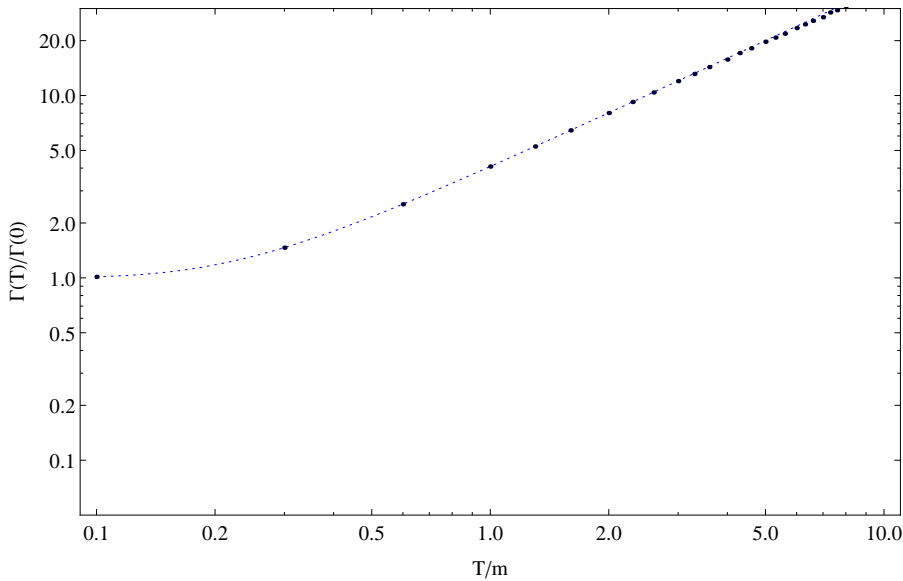}
    \caption{The rate $\Gamma$ in the model (\ref{L2}) as a function of $T$,
 normalised to its zero temperature value.
We set $\mathfrak{g}_1=\mathfrak{g}_2=0.1m$, $m_1=m_2=0.2 m$, $m_\xi=0.02 m$. 
The dashed light blue is the result one obtains when also neglecting thermal masses. 
The dark blue dots are numerical evaluations of (\ref{OffShell}), using the dispersion relations obtained from (\ref{disprelfinder}).  
\label{offshellplot2}}
\end{figure}

The non-monotonic behaviour shown in figure \ref{RePiPlot} makes it possible that (\ref{disprelfinder}) has more than one positive solution. The additional solutions could be interpreted as a collective excitation.
On the other hand, there is no guarantee that (\ref{disprelfinder}) has a solution at all because the principal value tends to be negative for small $\p$. This suggests that for some parameter choices there exists a minimum momentum for quasiparticles. For smaller momenta the spectral density (\ref{spectralfunction2}) is smooth, with no poles or narrow Breit-Wigner peaks, and there is no propagating excitation with a quasiparticle interpretation. Physically this would mean that the thermal damping in the plasma prevents excitations with small momenta from propagating; the resonances are so broad that their lifetime is comparable or shorter than their frequency.
Of course, the non-existence of a solution for (\ref{disprelfinder}) for a given mode $\p$ does not necessarily imply that there can be no quasiparticle with that momentum: (\ref{onepolerho}) also exhibits a pronounced peak if the LHS of (\ref{disprelfinder}) is sufficiently small.
\footnote{There appears to be no quasiparticle if the LHS of (\ref{disprelfinder}) is small, but non-zero, for $|p_0|<2m_\xi$ because in this region (\ref{onepolerho}) vanishes, and only exact zeros of the denominator seem to give a finite contribution to the spectral density (\ref{onepolerho}) in the limit $\epsilon\rightarrow 0$. However, this statement has to be taken with care because the function $\Pi_i^-$ in the numerator of (\ref{onepolerho})  for $p_0<2m_\xi$ only vanishes in the approximation (\ref{xiIntSelfEn}) at one-loop level. The full $\Pi_i^-$ is non-zero due to Landau damping.
In our numerical analysis we used the one-loop result (\ref{xiIntSelfEn}) for simplicity.}
Whether or not a well-defined resonance can be identified depends on the value of the LHS of (\ref{disprelfinder}) at its minimum.\footnote{Note that in this case (\ref{GammaFormula}) and (\ref{disprelfinder}) cannot be used to compute the width and precise dispersion relation, but the denominator of (\ref{spectralfunction2}) has to be evaluated in detail.} 
We restrict the analysis to the case where (\ref{disprelfinder}) has exactly one solution, for which the hierarchy $\Gamma_i\ll\Omega_i$ holds and allows to define a (possibly finite width) quasiparticle. Then (\ref{OffShell}) can be used to calculate $\Gamma$.
As a cross-check for the validity of (\ref{OffShell}), we evaluated the sum rule
\begin{equation}\label{sumrule}
\int\frac{d p_0}{2\pi} p_0\rho_i(p_0)=1
\end{equation}
for each point in the integration volume for each temperature.
It directly follows from the commutation relations $[\chi(x_1),\dot{\chi}(x_2)]|_{t_1=t_2}=i\delta(x_1-x_2)$ for a scalar field.
Using the sum rule we find that the approximations (\ref{BW}) and/or (\ref{onepolerho}) can fail at high temperature ($T\gg m_i,m$) for very hard ($|\p|\gg m_i$) and very soft ($|\p|\ll m_i$) momenta.  
This can easily be understood qualitatively. For soft momenta and $T\gg m_i$ the principal value in (\ref{disprelfinder}) dominates over $m_i$. As discussed before, in this regime the pole structure can be complicated and (\ref{onepolerho}), (\ref{BW}) may not provide useful approximations. 
For hard momenta one can in good approximation neglect all vacuum masses.
The thermal correction is small compared to $T$, hence the pole of $\rho_i$ and the spike-feature in ${\rm Re}\Pi^-$ (cf. figure \ref{RePiPlot}) approach each other near $p_0\sim|\p|$. The underlying assumption of the Breit-Wigner approximation is that $\Pi^R$ does not change considerably across the quasiparticle peak and can be replaced by its value at $\Omega_i$ in loop integrals. But near the spike, ${\rm Re}\Pi^R_i$ as a function of $p_0$ is steep and the factor $\mathcal{Z}_i$ is considerably different from unity. Physically this means that the $p_0$-integral in (\ref{sumrule}) is not strongly dominated by the peak region and the continuum part of $\rho_i$ cannot be neglected. 
In the numerical results plotted in figure \ref{offshellplot2} we chose parameters such that the deviation from (\ref{sumrule}) is less than 15$\%$ in the regions that contribute significantly to the integral in (\ref{OffShell}).

The thermal mass shift depends on both, temperature and momentum. 
It can have either sign, dependent on the choice of $m_i$, $T$, $\mathfrak{g}_i$ and $\textbf{p}$. 
It is therefore impossible to define a unique temperature $T_c$ at which the decay $\phi\rightarrow\chi_1\chi_2$ becomes kinematically forbidden. It can be  forbidden for some $\chi_i$-modes, which have received a sufficiently large positive ``thermal mass'' at a given temperature, but is still allowed for other modes if these have received a smaller (possibly even negative) correction to the dispersion relation. We therefore always have to insert the full dispersion relation into (\ref{OffShell}) \footnote{In the derivation of  (\ref{OffShell}) we approximated $\frac{\partial\Omega_i}{\partial|\textbf{p}|}=\frac{|\textbf{p}|}{\Omega_i}$, hence it can be applied as long as the deviation of the dispersion relation from a parabola is not too big.}. 
Figure \ref{offshellplot2} shows the temperature dependence of $\Gamma$ as a function of $T$.
For simplicity, we neglected thermal corrections to $m_\xi$. We checked that for the parameters we plot, a generic thermal mass does not change the results qualitatively.
There is clearly no suppression due to thermal masses even for $T\gg m,m_i$.
This is relatively generic if one chooses $m_i$ and $\mathfrak{g}_i$ much smaller than $m$.  
The reason is that the decay products have momenta $\lesssim m/2$. 
For $m_i\ll m\ll T$, these are soft, i.e. much smaller than the typical momentum $\sim T$ in the plasma. For soft momenta the corrections to the $\chi_i$ dispersion relations are small compared to $m$ (or even negative). That means the dispersion relations for very soft $\chi_i$-modes are almost free of thermal corrections and for these modes there are no large thermal masses even at high temperature.\footnote{The possibility of a vanishing thermal mass has been studied in a different context in \cite{Nakkagawa:2011ci}, where indications were found that the thermal mass may vanish in strongly coupled gauge theories.} 
The high sensitivity of the reheating rate to the dispersion relations at soft momenta is due to the fact that the inflaton decays at rest, hence the momenta of the decay products are determined by $m$ and not of the scale $T$. This is in contrast to most other transport phenomena in thermal equilibrium, where most of the energy is stored in modes with hard momenta $\sim T$ which receive thermal mass corrections of order $\alpha T$.    

\section{Example III: reheating by Landau damping}\label{QuarticInteraction}
We now consider the case that the $\chi_i$ self interactions are the same as in (\ref{L1}), but their coupling to $\phi$ is different. 
We use the following model
\begin{eqnarray}\label{L3}
\mathcal{L}&=& 
\frac{1}{2}\partial_{\mu}\phi\partial^{\mu}\phi
-\frac{1}{2}m^{2}\phi^{2}
+ \sum_{i=1}^2 \left(
\frac{1}{2}\partial_{\mu}\chi_{i}\partial^{\mu}\chi_{i}
-\frac{1}{2}m_{i}^{2}\chi_{i}^{2}-\frac{\lambda_i}{4!}\chi_i^4
-\frac{h_{i}}{4!}\phi\chi_{i}^{3}\right)
+ \mathcal{L}_{\rm bath}.\nonumber\\
\end{eqnarray}
The leading order contributions from the quartic interactions to $\Pi^-$ come from diagrams of the type shown in figure \ref{diagrams}b) and c).
In analogy to (\ref{trilinear}), they can be calculated from
\begin{eqnarray}\label{RisingSun}
\Pi_{\textbf{q}}^{-}(\omega)&=&2i\sum_i\frac{h_{i}^{2}}{6}\int\frac{d^{4}p}{(2\pi)^{4}}\frac{d^{4}k}{(2\pi)^{4}}\frac{d^{4}l}{(2\pi)^{4}}(2\pi)^{4}\delta^{(4)}(q-p-k-l) \rho_{i \textbf{p}}(p_{0})\rho_{i \textbf{k}}(k_{0})\rho_{i \textbf{l}}(l_{0})\nonumber\\
&&\times\Big[\big(1+f_B(p_0)\big)\big(1+f_B(k_0)\big)\big(1+f_B(l_0)\big)-f_B(p_0)f_B(k_0)f_B(l_0)\Big]
.
\end{eqnarray}
The integrand is again given by a product of spectral densities, weighted with distribution functions.
In the on-shell limit (\ref{rhofree}) it reads
\begin{eqnarray}\label{risingsun}
\lefteqn{\Pi^{-}_{\q}(\omega)=i\frac{h_{i}^{2}}{3}\int\frac{d^{3}\textbf{p}d^{3}\textbf{k}d^{3}\textbf{l}}{(2\pi)^{9}}(2\pi)^{3}\delta^{(3)}(\textbf{p}+\textbf{k}+\textbf{l}-\textbf{q})\frac{\pi}{8\Omega_{i \textbf{p}}\Omega_{i \textbf{k}}\Omega_{i \textbf{l}}}}\nonumber\\
&\times&\Big[\big(\left(1+f_{\textbf{p}}\right)\left(1+f_{\textbf{k}}\right)\left(1+f_{\textbf{l}}\right)-f_{\textbf{p}}f_{\textbf{k}}f_{\textbf{l}}\big)\nonumber\\
&&\phantom{\left(1+f_{\textbf{k}}\right)\left(1+f_{\textbf{l}}\right)}\left(\delta(\omega-\Omega_{i \textbf{p}}-\Omega_{i \textbf{k}}-\Omega_{i \textbf{l}})-\delta(\omega+\Omega_{i \textbf{p}}+\Omega_{i \textbf{k}}+\Omega_{i \textbf{l}})\right)\nonumber\\
&& + 3\big(f_{\textbf{p}}\left(1+f_{\textbf{k}}\right)\left(1+f_{\textbf{l}}\right)-\left(1+f_{\textbf{p}}\right)f_{\textbf{k}}f_{\textbf{l}}\big)\nonumber\\
&&\phantom{\left(1+f_{\textbf{k}}\right)\left(1+f_{\textbf{l}}\right)}\left(\delta(\omega+\Omega_{i \textbf{p}}-\Omega_{i \textbf{k}}-\Omega_{i \textbf{l}})-\delta(\omega-\Omega_{i \textbf{p}}+\Omega_{i \textbf{k}}+\Omega_{i \textbf{l}})\right)\Big],
\end{eqnarray}
with $f_{\textbf{p}}=f_B(\Omega_{i \textbf{p}})$ etc.
Again the integral is only non-vanishing when the submanifolds defined by the on-shell $\delta$-functions intersect somewhere in the integration volume. 
In vacuum, this would only be possible as long as the decay $\phi\rightarrow\chi_i\chi_i\chi_i$ is kinematically allowed, i.e. for $M>3M_i$. However, at finite temperature (\ref{RisingSun}) also includes terms that can be interpreted as scatterings $\phi\chi_i\leftrightarrow\chi_i\chi_i$. They give a contribution to $\Gamma$ that allows to reheat the plasma when the decay $\phi\rightarrow\chi_i\chi_i\chi_i$ is forbidden. 
In contrast to the trilinear interaction discussed in the previous section, these processes do not require an intermediate particle. 
This is a generic feature of vertices that connect more than three propagators.

\begin{figure}
  \centering
    \includegraphics[width=12cm]{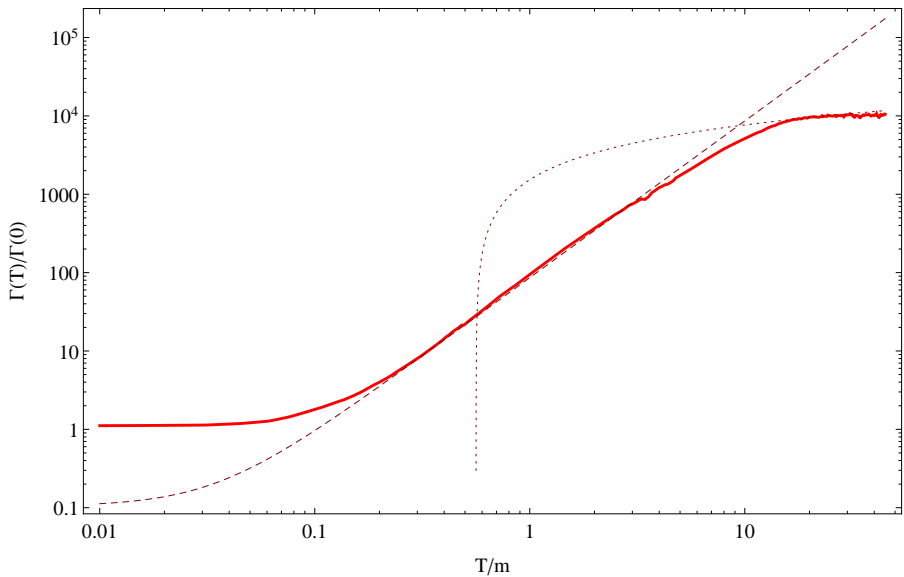}
    \caption{The contribution to $\Gamma$ from a $h_i/(4!)\phi\chi_i^3$-interaction for $\lambda_i=0.1$. The solid line is a numerical evaluation of (\ref{risingsun}), the dashed line the approximation (\ref{SimpleScatterings}) and the dotted line the approximation (\ref{quarticHighT}). All contributions are normalised to the value at $T=0$. For larger temperatures than plotted here our numerical evaluation is not reliable, but $\Gamma$ may decrease again due to the shrinking phase space for $\phi\chi_i\leftrightarrow\chi_i\chi_i$ scatterings. Note also that (\ref{risingsun}) does not take into account finite widths or resummed vertices.
\label{quarticplot}}
\end{figure}

However, since scatterings require $\chi_i$ quanta in the initial state, they are suppressed at low temperatures, when there are not sufficiently many scattering partners. 
In the high temperature regime (\ref{risingsun}) can be calculated analytically.
For our purpose, the most relevant case is $T\sim T_c$.
In \cite{Parwani:1991gq,Aarts:1996qi,Buchmuller:1997yw} an analytic expression has been obtained for $M=M_i$, \footnote{Note that the definition of $\Gamma$ varies by a factor $2$ in different articles.}  
\begin{equation}\label{SimpleScatterings}
\Gamma\simeq\sum_{i}\frac{h_{i}^{2}T^{2}}{768\pi M} \ \phantom{X} \ {\rm for} \ M=M_i 
.\end{equation}
This result should be approximately correct in the regime $T\simeq T_c$, where $\omega=M=2M_i$.
It is also interesting to consider the case $M_i\gg M$. In this regime we find, to leading order in $M/M_i$, 
\begin{equation}\label{quarticHighT}
\Gamma\simeq \frac{h_i^2 M}{6 (2\pi)^4}\frac{T^2}{M_i^2}\left(1+\log\left(\frac{81}{8}\frac{M_i}{M}\right)\right) \ \phantom{X} \ {\rm for} \ M\ll M_i.
\end{equation}
In figure \ref{quarticplot} we plot $\Gamma$ as a function of $T$ from numerical evaluation of (\ref{risingsun}). The analytic results (\ref{SimpleScatterings}) and (\ref{quarticHighT}) provide rather accurate approximations in some temperature regimes.
(\ref{risingsun}) only takes into account tree level processes $\phi\leftrightarrow\chi_i\chi_i\chi_i$ and $\phi\chi_i\leftrightarrow\chi_i\chi_i$ with all involved quasiparticles on-shell. 
In order to incorporate processes with more particles in the initial and final state as well as intermediate off-shell quasiparticles, one would have to include the finite widths, which are neglected in (\ref{rhofree}), and use resummed vertices. 

\begin{figure}
  \centering
    \includegraphics[width=12cm]{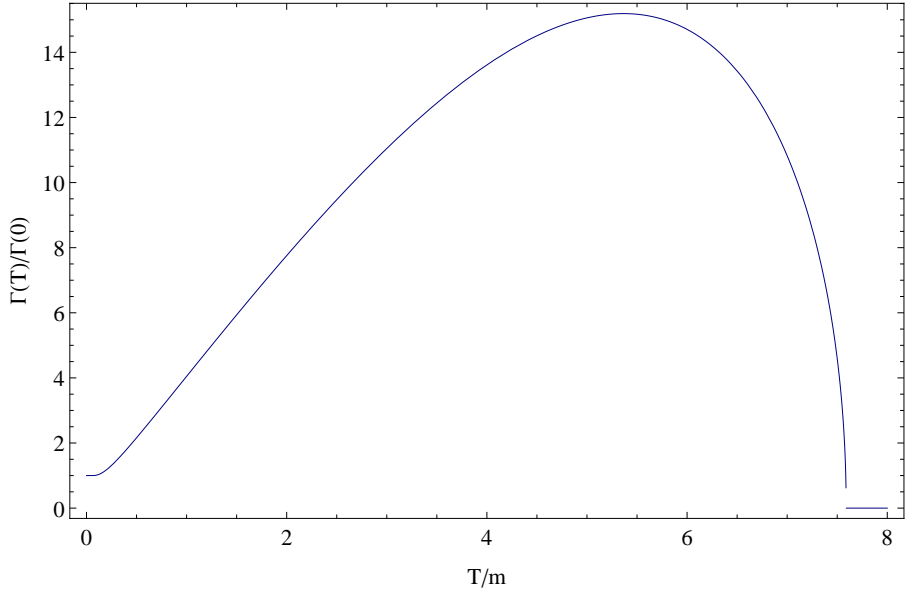}
    \caption{The decay rate (\ref{decayrate}) for $m_1=m_2=0.1m$ and $\lambda_1=\lambda_2=0.1$. \label{decayrateplot}}
\end{figure}

Let us combine the models (\ref{L1}) and (\ref{L3}). We assume that $\phi$ and $\chi_i$ couple by the trilinear $g\phi\chi_1\chi_2$ as well as the quartic $h_i/(4!) \phi\chi_i^3$ interactions. 
We compare the efficiency of decays $\phi\leftrightarrow\chi_1\chi_2$ to Landau damping by scatterings $\phi\chi_i\leftrightarrow\chi_i\chi_i$ in the regime $T\leq T_c$, both on-shell.\footnote{The discussion closely follows \cite{Drewes:2010pf}.}
We consider the simplest case, with $m_{1}=m_{2}=m_{\chi}$, $h_{1}=h_{2}=h$ and $\lambda_{1}=\lambda_{2}=\lambda$. 
The rate (\ref{dampingformel}) then reduces to
\begin{equation}\label{decayrate}
\Gamma_{{\rm decay}}=\frac{g^{2}}{16\pi M}\left[1-\left(\frac{2M_{\chi}}{M}\right)^{2}\right]^{1/2}\big(1+2f_{B}(M/2)\big)\theta(M-2M_{\chi}).\nonumber
\end{equation} 
The rate for $\phi\chi_i\leftrightarrow\chi_i\chi_i$ scatterings can be estimated by (\ref{SimpleScatterings}). 
At $T_{c}\approx(6(m^{2}-4m_{\chi}^{2})/\lambda)^{1/2}$  it takes the value
\begin{equation}
\Gamma_{{\rm scatter}}(T_{c})\approx2\frac{h^{2}m}{128\pi \lambda}\bigg(1-\Big(\frac{2m_{\chi}}{m}\Big)^{2}\bigg)
.\end{equation}
This can be compared to the decay rate (\ref{decayrate}) at the temperature where it is maximal and at $T=0$. 
The temperature dependence of (\ref{decayrate}) is
plotted in figure \ref{decayrateplot},
below $T_c$ it is very similar to the $M_1\neq M_2$ case discussed in section \ref{numerical}. 
For $0\leq T\leq T_c$ $\phi$ dissipates its energy via the decay $\phi\rightarrow\chi_1\chi_2$. 
Since both final states are bosonic, Bose enhancement causes $\Gamma$ to increase as a function of $T$. At the same time, the phase space for this process shrinks due to the large thermal $\chi_i$ masses.
Competition between these two effects determines the temperature $T_{max}$, at which $\Gamma$ is maximal. 
The value of $T_{max}$ can be found using the requirement $\partial\Gamma_{\rm decay}/\partial T=0$ at $T=T_{max}$, which allows to formulate the condition
\begin{equation}\label{TmaxCond}
m\left(6((2m_{\chi})^{2}-m^{2})+T_{max}^{2}\lambda\right)+T_{max}^{3}\lambda{\rm sinh}(m/T_{max})=0.
\end{equation}
For $T_{max}\gg m$ the ${\rm sinh}$ can be Taylor expanded to obtain 
\begin{equation}\label{Tmax1}
T_{max}\approx \left(\frac{3}{\lambda}\left(m^{2}-(2m_{\chi})^{2}\right)\right)^{1/2}
.\end{equation} 
With $M_{\chi}^{2}(T_{max})\approx(m_{\chi}^{2}+(m/2)^{2})/2$ this allows to estimate
\begin{equation}
\Gamma_{\rm decay}(T_{max})\approx \frac{g^{2}}{8\pi m}\sqrt{\frac{6}{\lambda}}\bigg(1-\Big(\frac{2m_{\chi}}{m}\Big)^{2}\bigg)
,\end{equation} 
where $T_{max}\gg m$ allowed to expand $f_{B}(m)$ in $m/T$. 
Comparison yields
\begin{eqnarray}
\frac{\Gamma_{\rm decay}(T_{max})}{\Gamma_{\rm decay}(T=0)}&=&\sqrt{\frac{24}{\lambda}}\bigg(1-\Big(\frac{2m_{\chi}}{m}\Big)^{2}\bigg)^{1/2}\\
\frac{\Gamma_{\rm scatter}(T_{c})}{\Gamma_{\rm decay}(T=0)}&=&\left(\frac{h}{2}\frac{m}{g}\right)^{2}\frac{1}{\lambda}\bigg(1-\Big(\frac{2m_{\chi}}{m}\Big)^{2}\bigg)^{1/2}\\
\frac{\Gamma_{\rm scatter}(T_{c})}{\Gamma_{\rm decay}(T_{max})}&=&\left(\frac{h}{2}\frac{m}{g}\right)^{2}\sqrt{\frac{1}{24\lambda}}=\sqrt{\frac{\lambda}{24}}\frac{\Gamma_{\rm scatter}(T_{c})}{\Gamma_{\rm decay}(T=0)}\Big|_{m_{\chi}=0}\label{ScatterOverDecaymax}
.\end{eqnarray}
Remarkably (\ref{ScatterOverDecaymax}) does not depend on $m_{\chi}$. 
In the interesting case $m_{\chi}\ll m$ also the other ratios are very simple. 
The parameter that governs the maximal amount of amplification of the decay rate by Bose enhancement is $\sim(24/\lambda)^{1/2}$. 
The parameter that determines whether relaxation via scatterings is efficient at $T_{c}$ is $(h^{2}/\lambda)\cdot(m/g)^{2}$. For $g/m\sim h$ the rate $\Gamma_{\rm scatter}(T_{c})$ can easily be bigger than $\Gamma_{\rm decay}(T=0)$ or even $\Gamma_{\rm decay}(T_{max})$.
That means that scatterings can heat the universe to temperatures above $T_c$. 

\section{Example IV: production of fermions}\label{yukawasection}
Matter in the real world is composed of fermions with gauge interactions. 
So far we have only considered the case that $\phi$ first dumps its energy into other bosons and assumed that fermions get created in the subsequent decay chain or inelastic scatterings in the plasma. This is justified because the transition into bosonic final states is usually Bose-enhanced, while those into fermionic final states are Pauli suppressed. In this last example we study $\Gamma$ in a model where $\phi$ directly couples to fermions, 
\begin{eqnarray}\label{L4}
\mathcal{L}&=& 
\frac{1}{2}\partial_{\mu}\phi\partial^{\mu}\phi
-\frac{1}{2}m^{2}\phi^{2}
+\bar{\Psi}\left(i\Slash{\partial}-\fermionmass{m}\right)\Psi
-Y\phi\bar{\Psi}\Psi
-\upalpha\bar{\Psi}\gamma^\mu A_\mu\Psi 
-\frac{1}{4}F_{\mu\nu}F^{\mu\nu}
+ \mathcal{L}_{\rm bath}.\nonumber\\
\end{eqnarray}
Here $A_\mu$ is a gauge field and $F_{\mu\nu}$ is the field strength tensor.
The rate $\Gamma$ calculated from the diagram shown in figure \ref{diagrams}d) is given by
\begin{eqnarray}\label{FermionLoopintegral}
\lefteqn{\Pi^{-}_{\textbf{q}}(\omega)=-iY^{2}\int\frac{d^{4}p}{(2\pi)^{4}}
{\rm tr}\left(\uprho_{\textbf{p}}(p_{0})\uprho_{\textbf{p}-\textbf{q}}(p_0-\omega)\right)}\\
&&\times\Big(\big(1-f_{F}(p_{0})\big)\big(1-f_{F}(\omega-p_0)\big)-f_F(p_0)f_F(\omega-p_0)\Big)\nonumber
,\end{eqnarray}
where $f_F(\omega)=(e^{\omega/T}+1)^{-1}$ is the Fermi-Dirac distribution.
\subsection{Analytic results for temperatures $T<T_c$}
\begin{figure}
  \centering
    \includegraphics[width=12cm]{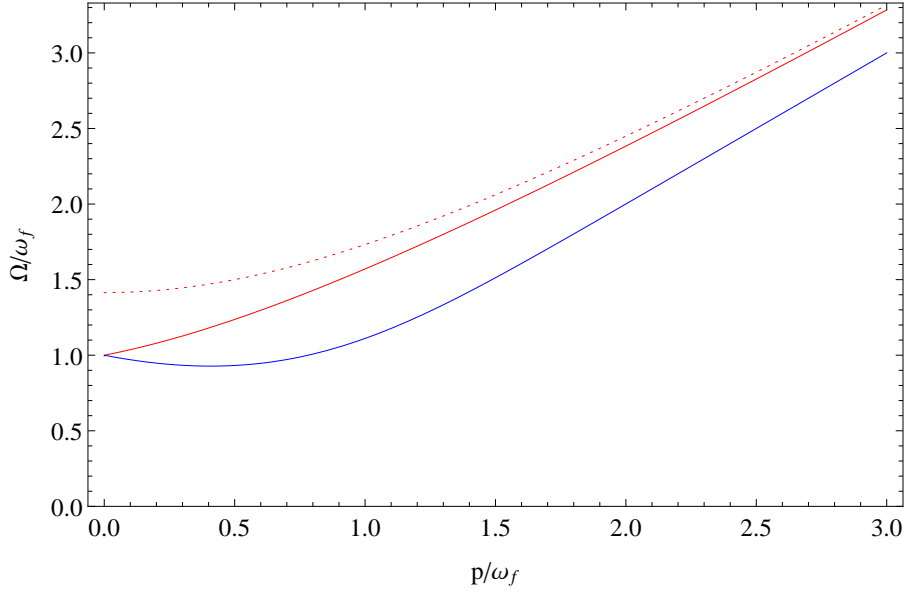}
    \caption{The fermionic dispersion relations $\Upomega_+$ (red line) and $\Upomega_-$ (blue line) in units of the plasma frequency $\upomega_f=\fermionmass{M}_f/\sqrt{2}$. The dashed line shows the approximation $(\textbf{p}^2+\fermionmass{M}_f^2)^{1/2}$ for hard momenta. 
For simplicity we assumed in (\ref{L4}) that $\Psi$ is charged under an $U(1)$ gauge group; the behaviour in a non-Abelian plasma is similar. \label{disprelations}}
\end{figure}

For $T<\fermionmass{m}$ thermal corrections to the fermion properties are negligible and we can use the bare fermion spectral density to evaluate (\ref{FermionLoopintegral})
\begin{equation}
\uprho_{\textbf{p}}^{\rm free}(p_{0})=2\pi {\rm sign}(p_0) (\slashed{p}+\fermionmass{m})\delta(p^2-\fermionmass{m}^2)
.\end{equation} 
The result is
\begin{equation}\label{FermionGammaSimpleI}
\Gamma=\frac{Y^2}{8\pi}m\left[1-\left(\frac{2\fermionmass{m}}{m}\right)^2\right]^{3/2}\left[1-2f_F(m/2)\right]\theta(m-2\fermionmass{m}) \ \hspace{0.5cm} {\rm for} \ T<\fermionmass{m}
\end{equation}
In the more interesting regime $T\gg \fermionmass{m}$ (we neglect $\fermionmass{m}$ in the following) we again have to use resummed spectral densities as in the scalar case. 
In the hard thermal loop (HTL) approximation one obtains \cite{Weldon:1982bn}
\begin{equation}
\uprho_{\textbf{p}}(p_{0})=\frac{1}{2}\left((\gamma_{0}-\hat{\bf{p}}\pmb{\gamma})\uprho_{+}+(\gamma_{0}+\hat{\bf{p}}\pmb{\gamma})\uprho_{-}\right)\label{HTLrho}
.\end{equation}
Here $\hat{\bf{p}}\pmb{\gamma}=p_{i}\gamma_{i}/|\textbf{p}|$. The functions 
\begin{equation}
\uprho_{\pm}(p)\simeq2\pi[\uprho_{\pm}^{\rm pole}(p)+\uprho^{\rm cont}_{\pm}(p)]
\end{equation}
are the sum of singular contributions $\uprho_{\pm}^{\rm pole}$ and continua $\uprho^{\rm cont}_{\pm}$.
The poles of the singular parts define quasiparticles with energies $\Upomega_{\pm}$, 
\begin{equation}\label{polepart}
\uprho_{\pm}^{\rm pole}(p)= Z_{\pm}\delta(p_{0}-\Upomega_{\pm})+Z_{\mp}\delta(p_{0}+\Upomega_{\mp}),
\end{equation}
the continuous part is given by
\begin{eqnarray}
\lefteqn{\uprho^{\rm cont}_{\pm}(p)=\theta(1-x^{2})\frac{y^{2}}{|\textbf{p}|}(1\mp x)}\nonumber\\
&\times&\Bigg[\Bigg(1\mp x\pm y^{2}\left((1\mp x)\ln\left|\frac{x+1}{x-1}\right|\pm 2\right)\Bigg)^{2}
+\pi^{2}y^{4}(1\mp x)^{2}\Bigg]^{-1}.
\end{eqnarray}
We used the notations $x=p_{0}/|\textbf{p}|$ and $y=\frac{1}{2}\fermionmass{M}_{f}/|\textbf{p}|$. 
The thermal mass reads $\fermionmass{M}_{f}=\upalpha T/2$. $\fermionmass{M}_{f}$ is sometimes referred to as {\it asymptotic mass}, it differs by a factor $\sqrt{2}$ from the plasma frequency $\upomega_f$ at $|\textbf{p}|=0$.
The residues are
\begin{equation}
Z_{\pm}=\frac{\Upomega_{\pm}^{2}-\textbf{p}^{2}}{4y^{2}\textbf{p}^{2}}
.\end{equation}
The dispersion relations $\Upomega_{+}$ and $-\Upomega_{-}$ are the solutions to
\begin{equation}
0=p_{0}-|\textbf{p}|\left[1+y^2\left((1-x)\ln\frac{x+1}{x-1}+2\right)\right]
.\end{equation}
There are two types of quasiparticles, with dispersion relations $\Upomega_+$ and $\Upomega_-$.  The former can be interpreted as one-particle states that are screened by the plasma, the latter are collective resonances often dubbed ``holes'' or ``plasminos''.
This complicated structure makes it obvious that correct results for fermions cannot be obtained when simply replacing intrinsic by thermal masses.
The dispersion relations $\Upomega_\pm$ can be expressed in terms of the Lambert $W$-function as \cite{Kiessig:2010pr}
\begin{equation}
\Upomega_+=|\textbf{p}|\frac{W_{-1}(s)-1}{W_{-1}(s)+1} \ , \ \Upomega_-=-|\textbf{p}|\frac{W_{0}(s)-1}{W_{0}(s)+1}
\end{equation}
with $s=e^{-(y+1)}$. They are plotted in figure \ref{disprelations}.

At temperatures $T\ll M/\upalpha$ we can approximate (\ref{FermionLoopintegral}) by neglecting $\uprho_\pm^{\rm cont}$.
Since $m\gg\fermionmass{m}$, the daughter particles have hard momenta $|\textbf{p}|\gg \fermionmass{m}, \fermionmass{M}_f$. 
In this regime \footnote{In principal the HTL approximation only holds for soft momenta $|\textbf{p}|\ll T$. We here use it for arbitrarily hard momenta, assuming that medium related modifications of the spectral density play no role for $|\textbf{p}|\gtrsim T$.} one can approximate 
\begin{eqnarray}
\begin{tabular}{c c c}
$Z_{+}\simeq 1+y^{2}\left(1+2\ln(y)\right)$ & & $Z_{-}\simeq \upalpha^{-1}y^{-2}\exp\left(-y^{-2}\right)$\label{Residuen}
\end{tabular}
\end{eqnarray}
and 
\begin{eqnarray}
\begin{tabular}{c c c}
$\Upomega_{+}\simeq |\textbf{p}|(1+2y^{2})\simeq\left(\textbf{p}^{2}+\fermionmass{M}_{f}^{2}\right)^{1/2}$ & & $\Upomega_{-}\simeq|\textbf{p}|(1+2\upalpha^{-1}\exp(-y^{-2}))$\label{FermionPoles}
\end{tabular}
\end{eqnarray}
The dispersion relation $\Upomega_{+}$ for the screened particle approaches that of a particle with mass $\sim\fermionmass{M}_{f}=\upalpha T/2$. This thermal mass only appears in the dispersion relation and does not break the chiral symmetry \cite{Weldon:1982bn}, as one can see from (\ref{HTLrho}). 
The holes effectively become massless for hard momenta $\gg \fermionmass{M}_f$. 
However, the residue $Z_{-}$ is exponentially suppressed for large momenta. The physical reason is that holes are collective phenomena. Such excitations appear at length scales
  which are associated with {\it soft} momentum modes $\sim \upalpha T$ 
  and
longer than the typical inter-particle distance $\sim 1/T$.
The occupation numbers of soft modes $\sim \upalpha T$ in a thermal plasma are small compared to the hard modes $\sim T$. Therefore it is usually argued that processes involving holes do not contribute significantly to the damping unless other channels are forbidden or they are resonantly amplified\footnote{Similar arguments apply to hydrodynamic and ultrasoft modes studied in \cite{Hidaka:2011rz}.}. 
Thus, the spectral density can be approximated as
\begin{equation}\label{rhoFermiQED}
\uprho_{\textbf{p}}(p_{0})\simeq \pi\left((\gamma_{0}-\hat{\textbf{p}}\pmb{\gamma})\delta(p_{0}-\Upomega_{+})+(\gamma_{0}+\hat{\textbf{p}}\pmb{\gamma})\delta(p_{0}+\Upomega_{+})\right)
.\end{equation}
Insertion of (\ref{rhoFermiQED}) into (\ref{FermionLoopintegral}) yields for $|\textbf{q}|=0$
\begin{eqnarray}\label{FermionGammaSimpleII}
\Gamma=\frac{Y^{2}}{8\pi}M\left[1-\left(\frac{2\fermionmass{M}_{f}}{M}\right)^2\right]^{1/2}\left(1-2f_{F}(M/2)\right)\theta(M-2\fermionmass{M}_{f}) \ \hspace{0.5cm} {\rm for} \ \fermionmass{m}\ll \fermionmass{M}_f < M
\end{eqnarray}
where we approximate $\Upomega_{+}\approx(\textbf{p}^{2}+\fermionmass{M}_{f}^{2})^{1/2}$.
This result can be used to formally define $T_c$ as the temperature where $M=2\fermionmass{M}_{f}$. 
The analytic formula (\ref{FermionGammaSimpleII}) for small $\upalpha$ is in good agreement with the numerical results found in \cite{Enqvist:2004pr} and our own results shown in figure \ref{FermionFullFig}. 
Note that (\ref{FermionGammaSimpleII}) cannot be obtained from (\ref{FermionGammaSimpleI}) by simply replacing the bare mass $\fermionmass{m}$ with the thermal mass $\fermionmass{M}_f$.
We can draw two important conclusions in the regime $\fermionmass{M}_f < m<T$. First, if the $\phi$-couplings to fermions and bosons are of comparable strength, then
comparison with (\ref{decayrate}) shows that the contributions to $\Gamma$ from decay into fermions is suppressed due to Pauli blocking. Second, the effect of the thermal mass $\fermionmass{M}_f$ on $\Gamma$ is relevant only for $T\gtrsim m/\upalpha$, while Pauli suppression is effective already at $T\gtrsim m$.

\subsection{Numerical results for all temperatures}
\begin{figure}
  \centering
    \includegraphics[width=6cm]{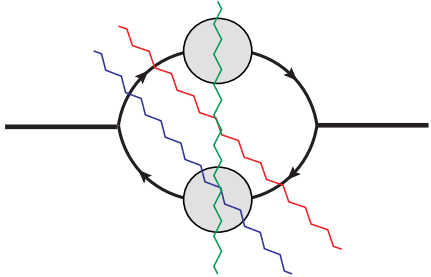}
    \caption{Cuts through the $\phi$ self-energy with resummed fermion propagators.
\label{fermioncuts}}
\end{figure}

\begin{figure}
  \centering
    \includegraphics[width=6cm]{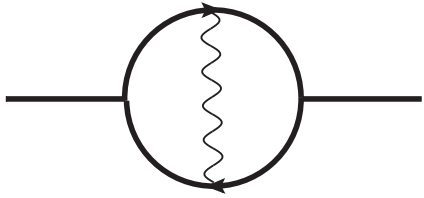}
    \caption{A ladder diagram with one rung. For the Yukawa coupling considered here it has been argued that vertex corrections from such diagrams are sub-dominant \cite{Thoma:1994yw}. 
It has been found in \cite{Anisimov:2010gy} that resummation of ladder diagrams can give a large contribution to relaxation rates, cf. also \cite{Garbrecht:2013gd} in that context. However, in that work the momenta of typical external particles is $\sim T$.  Therefore the particles have collinear momenta in the rest frame of the bath and propagate parallel for a sufficiently long time to exchange several gauge bosons. It is precisely this collinear enhancement that leads to the large contribution. Furthermore, one of the internal lines in that calculation is a boson. We therefore expect that the effect of ladder diagrams is smaller in the case of reheating and does not change our findings qualitatively.    
    \label{ladders}}
\end{figure}

It is, however, not clear that the above conclusions hold at higher temperature.
The reason for the Pauli suppression for $T>m$ is that the energy of the $\Psi$-quasiparticles in $\phi\rightarrow\Psi\bar{\Psi}$ decays is $\sim m/2$, independently of the temperature.  
Cuts through the resummed propagators as shown in figure \ref{fermioncuts} include scatterings $\phi\Psi\leftrightarrow \Psi\gamma$. In these Landau-damping processes, the fermions can have hard momenta and energies $\sim T$, hence there is no Pauli blocking. 
The continuous parts $\uprho_\pm^{\rm cont}$ of the fermion spectral densities consistently incorporate the leading order contribution from Landau damping in the resummed propagators. They do, however, not include contributions from ``ladder diagrams'' as shown in figure \ref{ladders}. Inclusion of these would require not only resummed propagators, but also resummed vertices. In \cite{Karsch:2000gi} it was argued that corrections to the decay rate of a scalar into fermions with gauge interactions from resummation of vertices are of higher order, see also \cite{Thoma:1994yw}\footnote{Note that there is no Ward-Takahashi identity for the Yukawa vertex, as there would be for gauge interactions.}. 
For the current purpose we will assume that this statement is correct, though it has been found in other contexts that ladder diagrams can be relevant (cf. caption of figure \ref{ladders}).
This implies that evaluation of (\ref{FermionLoopintegral}) with the full resummed spectral densities (\ref{HTLrho}) is sufficient to include all processes at leading order in $\upalpha$.
We split the product of spectral densities in (\ref{FermionLoopintegral}) in three parts. 

\begin{figure}
  \centering
    \includegraphics[width=14cm]{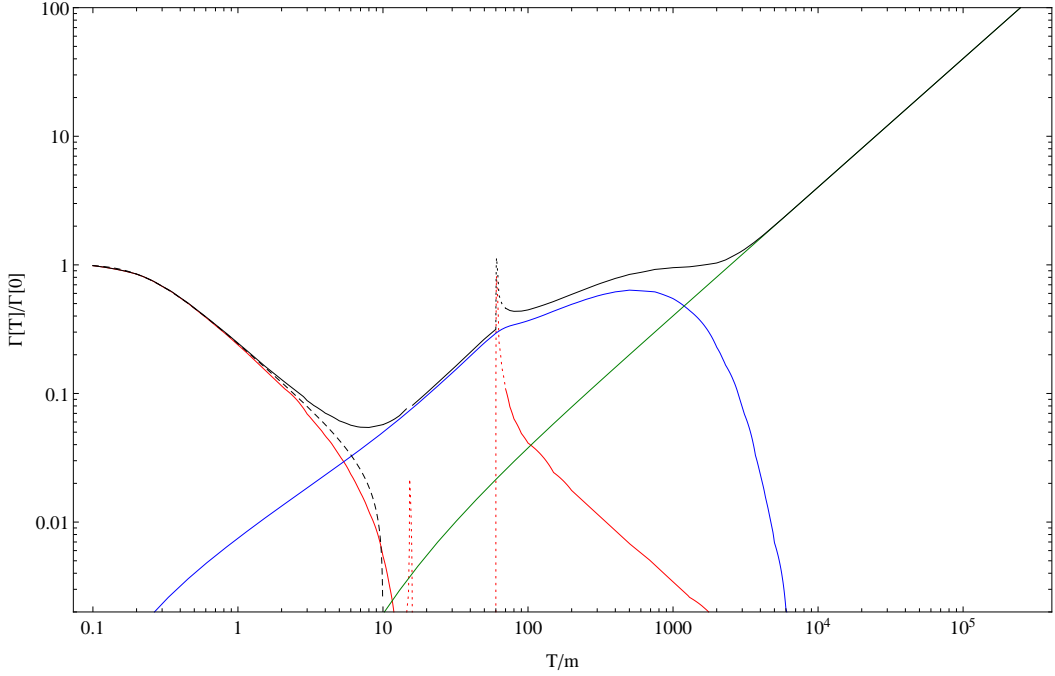}
    \caption{The solid black line shows the total $\Gamma$ in the model (\ref{L4}) calculated from (\ref{FermionLoopintegral}) and normalised to its $T=0$ value for $\upalpha=0.1$. 
The dashed black line is the approximation (\ref{FermionGammaSimpleII}).
The colourful lines show the contributions from the individual terms (\ref{A})-(\ref{C}). 
They can be identified with the cuts in figure \ref{fermioncuts} as indicated by the colour coding. 
The red line is the contribution from term (\ref{A}). In the regime $T<T_c\sim 10m$ it comes from the on-shell decays of $\phi$-particles into two $\Psi$-particles. The non-zero contribution for temperatures slightly above $T_c$ originates from on-shell decays of $\phi$-particles into two collective fermionic excitations. The contribution for $T>\tilde{T}_c\sim 60m$ originates from annihilations of a $\phi$-particle by a collective $\Psi$-excitations to form a  $\Psi$-particle. The contributions from processes involving collective excitations are divergent near the thresholds due to a van Hove singularity in the density of states, see discussion in the text. In these regions the finite widths of the quasiparticles have to be taken into account to obtain a physically meaningful and finite result, which goes beyond the HTL approximation (\ref{HTLrho}). We do not do that in this work and indicate the corresponding regions by dotted lines. The blue line is the contribution from term (\ref{B}), which includes $\phi\Psi\rightarrow\Psi\gamma$ scatterings and is non-vanishing in the regime $T_c<T<\tilde{T}_c$. The green line is the contribution from term (\ref{C}), which includes various scattering processes involving fermions and photons. They are not Pauli suppressed. At low temperature they are negligible because they contribute at higher order in $\upalpha$. At high temperatures, however, this suppression is overcome by the high density of scattering partners and the Bose-enhancement for photons.  
\label{FermionFullFig}}
\end{figure}

The first part is related to the red cut in figure \ref{fermioncuts} and contains only contributions from the on-shell parts $\uprho_\pm^{\rm pole}$ of the spectral densities. It reads
\begin{eqnarray}\label{A}
\lefteqn{\Pi^-_{\textbf{0}}(m)|_{ 
{\rm p}
\times{\rm p}
}
=-iY^2\frac{2}{\pi}\int d|\textbf{p}| \textbf{p}^2\Bigg[
Z_+^2  \big[1-f_F(\Upomega_+) -f_F(m-\Upomega_+)\big] \delta(m-2\Upomega_+)}\nonumber\\
&+&Z_+Z_-\big[1-f_F(\Upomega_+)-f_F(m-\Upomega_++f_F(\Upomega_-)-f_F(\Upomega_-+m)\big]\delta(m+\Upomega_--\Upomega_+)\nonumber\\
&+&Z_-^2\big[1-f_F(\Upomega_-)-f_F(m-\Upomega_-)\big]\delta(m-2\Upomega_-)\Bigg]
\end{eqnarray}
Since the finite widths are neglected in (\ref{polepart}) just as in (\ref{rhofree}) for the scalar case, evaluation of (\ref{FermionLoopintegral}) with (\ref{polepart}) leads to a qualitatively similar behaviour as in (\ref{dampingformel}); in particular, temperatures $T_c$ and $\tilde{T}_c$ can be defined. 
The $\delta$-function allows to evaluate the integral analytically, but the zeros of their arguments have to be found numerically. The contribution to $\Gamma$ from (\ref{A}) is shown as red line in figure \ref{FermionFullFig}.  The term (\ref{A}) contains three types of contributions. The terms $\propto \delta(m-2\Upomega_+)$ describe $\phi$-decays into dressed $\Psi$-particles. They clearly dominate at $T<T_c\sim m/\upalpha$, where (\ref{FermionGammaSimpleII}) provides an excellent approximation. Terms $\propto \delta(m-2\Upomega_-)$ describe the decay into two collective fermionic excitations. Since (\ref{Residuen}) shows that $Z_-$ is suppressed for hard momenta, they only contribute significantly near $T_c$, where the decay products' momenta are soft.  
Their main effect, visible in figure \ref{FermionFullFig}, is that (\ref{A}) is non-zero for temperatures slightly above $T_c$, where the decay into dressed particles is already kinematically forbidden ($2\Upomega_+>m$), but the decay into collective excitations is still allowed ($2\Upomega_-<m$).\footnote{A precise calculation including ladder diagrams in this regime, where the $\Psi$ and $\bar{\Psi}$ are non-relativistic, should reveal that a similar contribution for $T$ slightly above $T_c$ comes from the production of a $\Psi\bar{\Psi}$ bound state, which is slightly lighter than free $\Psi$ and $\bar{\Psi}$.} 
However, they also become kinematically forbidden at only slightly higher temperatures. Both of these contributions are suppressed by Pauli blocking at $T>m$. 
The contribution $\propto\delta(m+\Upomega_--\Upomega_+)$ comes from processes in which a $\phi$-particle and a collective $\Psi$ excitation form a $\Psi$-particle. These become kinematically allowed above a second critical temperature $\tilde{T}_c$, where $\Upomega_+>m+\Upomega_-$.\footnote{While for scalars it was necessary that there are two fields $\chi_i$ with $M_1\neq M_2$ for the similar processes $\phi\chi_i\leftrightarrow\chi_j$ to be kinematically allowed, they are always possible for fermions because the hole provides a second particle with different ``mass''.}  
The contribution from such processes is generally small, only near the threshold at $\tilde{T}_c$ it is amplified by a {\it van Hove singularity}. This singularity originates from the infinite density of states $(d\textbf{p})/(d\Upomega_-)$ at the local minimum of $\Upomega_-$ for $|\textbf{p}|\neq 0$ visible in figure \ref{disprelations}. For most parameter choices physical observables are not affected by the singularity because the density of states only diverges in one point and gives a finite contribution when convolved with smooth functions. Only if the mass shell defined by the zero of $\delta(m+\Upomega_--\Upomega_+)$ happens to coincide with the van Hove singularity $\Gamma$ is divergent. This happens at $\tilde{T}_c$. The divergence is, however, not physical and can be removed by taking into account the finite width of the quasiparticles, which is neglected in the hard thermal loop approximation (\ref{polepart}). This simply means that we cannot trust our approximations in a small temperature interval around $\tilde{T}_c$, which we exclude from the following considerations. Similar behaviour was found for the dilepton production rate in a quark-gluon plasma \cite{Braaten:1990wp}. 

The second term comes from cuts through the $\phi$-self-energy that cut through one $\Psi$-self-energy (blue in figure \ref{fermioncuts}) and reads
\begin{eqnarray}\label{B}
\Pi^-_{\textbf{0}}(m)|_{ 
{\rm p}
\times{\rm c}
}
&=&-iY^2\frac{4}{\pi}\int d|\textbf{p}| \textbf{p}^2\Bigg[ Z_+\bigg[\big(1-f_F(\Upomega_+) -f_F(m-\Upomega_+)\big)\uprho_-^{\rm cont}(\Upomega_+-m)\nonumber\\ 
&&+ \big(f_F(\Upomega_+) -f_F(\Upomega_++m)\big)\uprho_-^{\rm cont}(\Upomega_++m)\bigg]\nonumber\\
&+& Z_-\bigg[\big(1-f_F(\Upomega_-) -f_F(m-\Upomega_-)\big)\uprho_+^{\rm cont}(\Upomega_--m) \nonumber\\
&&+ \big(f_F(\Upomega_-) -f_F(\Upomega_-+m)\big)\uprho_+^{\rm cont}(\Upomega_-+m)\bigg]\Bigg]
\end{eqnarray}
The continuum parts $\uprho_\pm^{\rm cont}$ originate from the imaginary parts of the $\Psi$-self-energies, hence including them means that we take into account dissipative effects for $\Psi$ though the thermal width is neglected in (\ref{polepart}).
It includes processes such as $\phi\Psi\leftrightarrow\gamma\Psi$, with intermediate $\Psi$. 
These are kinematically always allowed, hence (\ref{B}) is nonzero for arbitrary temperatures there is no equivalent to $T_c$. However, they always contain fermions in the initial and final state. Since the momentum of the outgoing fermion kinematically must be smaller than that of the incoming one (because the $\phi$ is at rest), they tend to be Pauli suppressed. Indeed this contribution never exceeds the vacuum $\Gamma$ in figure \ref{FermionFullFig}.

The last term reads
\begin{eqnarray}\label{C}
\lefteqn{\Pi^-_{\textbf{0}}(m)|_{ 
{\rm c}
\times{\rm c}
}
=-iY^2\frac{2}{\pi}\int d|\textbf{p}| dp_0 \textbf{p}^2 \big[1-f_F(p_0)-f_F(m-p_0)\big]}\nonumber\\
&\times&\big[\uprho_+^{\rm cont}(p_0)\uprho_-^{\rm cont}(p_0-m)+\uprho_-^{\rm cont}(p_0)\uprho_+^{\rm cont}(p_0-m)\big]
\end{eqnarray}
It contains the contributions from cuts through the $\phi$-self-energy that cut through both $\Psi$-self-energies (green cut in figure \ref{fermioncuts}). They can have more particles in the final state. This increased flexibility in the phase space leaves more room for the fermions to have hard momenta and avoid Pauli suppression. At the same time, transitions with bosonic final states with soft momenta are enhanced. At low temperatures such processes are negligible because they are of higher order in the coupling, but at high temperature this suppression is cancelled by the high density of scattering partners (formally the behaviour of the Bose-Einstein distribution for soft modes). At temperatures $T\gg \tilde{T}_c$ the term (\ref{C}) becomes the dominant source of dissipation; it continues to grow with $T$ and can exceed the vacuum decay rate by orders of magnitude. 

While it is usually assumed that fermionic channels are not important for reheating because they are Pauli suppressed, our result suggests that they can give a huge (possibly dominant) contribution to $\Gamma$ at very large temperatures. If the Yukawa coupling is the only interaction that $\phi$ has, then the universe may never reach this temperature regime because of the suppression of $\Gamma$ around $T\lesssim m/\upalpha$.
But if some other interaction is efficient in this regime, then processes involving fermions can be crucial at higher temperatures.

\section{Conclusions}\label{Conclusions}
We calculated the relaxation rate $\Gamma$ of a scalar field $\phi$ in a plasma of scalars with trilinear and quartic couplings and fermions with gauge interaction, using resummed perturbation theory at finite temperature. 
We focused on the regime where a quasiparticle description applies and the dissipation is dominated by perturbative processes involving individual quanta rather than particle production from the time varying background field $\langle\phi\rangle$. 
We paid particular attention to the temperature dependence of the phase space, caused by the modified quasiparticle dispersion relations in the plasma, and to quantum statistical effects. 

Many transport phenomena can in good approximation be understood in the description of a hot plasma as a collection of quasiparticles with momentum independent thermal masses.
The reason is that, in weakly coupled gauge theories at temperature $T$ in an adiabatically changing background, the dispersion relations for quasiparticles with hard momenta $\p\sim T$ asymptotically approach $\omega\simeq(\p^2+M_i^2)^{1/2}$, where $M_i\sim\alpha_i T$ is a momentum independent thermal mass. Since most particles in a thermal plasma have momenta $\sim T$, this description is appropriate for most microphysical processes.

This is, however, not the case for the decay of heavy particles with mass $m$ at rest or the dissipation of the oscillating zero mode of a field, such as the inflaton. 
The reason is that in such a decay the typical momentum of the decay products is not $\sim T$, but of order $m/2$ or smaller. For $T\gg m$ such momenta are soft compared to typical energies $\sim T$ in the plasma. This makes $\Gamma$ sensitive to the dispersion relations for soft modes. They can be complicated functions of momentum that differ for different interactions.

$\Gamma$ is of crucial importance to understand the thermal history of the universe after inflation, as it yields the rate of perturbative reheating. Our results allowed to investigate the possibility that thermal masses impose an upper bound on the temperature in the early universe by kinematically blocking the inflaton decay into the corresponding particles.  
We conclude that thermal masses can have a considerable effect on the efficiency of perturbative reheating.
However, a strict bound on the reheating temperature from kinematic blocking of inflaton decay can only be obtained in very special setups.
In general, other mechanisms allow the inflaton to dissipate its energy into the primordial plasma at high temperature. These in particular include Landau damping by scattering. Even when large thermal masses require some intermediate particle to be off-shell, these can be relevant due to Bose or collinear enhancement. 

A main conclusion of our study is that,
even if one is only concerned with perturbative reheating, 
it is usually not possible to express the temperature at the onset of the radiation dominated era or the maximal temperature in the early universe in terms of the parameters appearing in the Lagrangian in a simple way. 
The reason is that the temperature dependence of the dissipation rate $\Gamma$ 
can be highly nontrivial. The shape of the function $\Gamma(T)$ strongly depends on the interactions of the fields within the primordial plasma amongst each other, and not only on the couplings of the inflaton. This is because these interactions are involved when $\phi$ dissipates energy via scatterings, and they also modify the phase space via the dispersion relations in the plasma .
Reheating is not an instantaneous process, but takes place over a time interval $\sim1/\Gamma$ during which the universe expands. Hence, the temperature dependence of $\Gamma$ is crucial to determine the time evolution of the effective temperature in the plasma.  
If reheating is only (or mainly) driven by perturbative processes, then 
the ``naive'' reheating temperature defined in (\ref{naive}) acts as upper bound for $T$. The real temperature is always lower because the universe expands during the reheating process. 

In order to determine the time evolution of $T$, including its maximal value and its value at the onset of the radiation dominated era, it is crucial to take into consideration the effect the medium has on $\Gamma$.
The results found in this work are suitable for a quantitative description in periods during which the dominant contribution to $\Gamma$ comes from perturbative processes. 
Even if non-perturbative particle production and a parametric resonance dominate during an early preheating phase, perturbative dissipation described by $\Gamma$ usually dominates during the late phase of reheating. Furthermore, perturbative processes can be crucial during a preheating phase if they lead to instant or combined preheating \cite{Felder:1998vq,Micha:2004bv,GarciaBellido:2008ab,Mukaida:2012bz} by allowing the produced particles to decay efficiently within one $\phi$-oscillation \cite{Mukaida:2012qn,Mukaida:2012bz}. 
In addition, $\Gamma$ also determines the rate of dissipation during inflation itself \cite{BasteroGil:2010pb}, which is crucial in models of warm inflation \cite{Berera:1995ie}.

The effects we describe are relevant beyond the domain of inflation if $\phi$ is identified with another scalar quantity, e.g. order parameter during a phase transition. 
The evolution of other flat directions in field space in the early universe can be affected \cite{Yokoyama:2004pf,Enqvist:2011pt,Sanchez:2012tk,Yokoyama:2006wt}. 
Apart from instant or combined preheating,  other phenomena in which perturbative and non-perturbative processes work together, e.g. curvaton decay \cite{Enqvist:2012tc} or Affleck-Dine baryogenesis \cite{Allahverdi:2000zd}, can be affected.
 
Finally, our results may also be important for other transport phenomena in cosmology where thermal effects have been found to be relevant, including leptogenesis \cite{Giudice:2003jh,Akhmedov:1998qx,Asaka:2005pn,Kiessig:2010pr,Anisimov:2010aq,Anisimov:2010gy,Laine:2011pq,Garbrecht:2011aw,Frossard:2012pc,Garbrecht:2013gd}, the production of dark matter \cite{Hamaguchi:2011jy,Saikawa:2012uk}, warm baryogenesis \cite{BasteroGil:2011cx} or
axion production \cite{Masso:2002np}\\ \\ 
\newline
\vspace{0.5cm}
{\large \textbf{Acknowledgements}} -
We thank Fernando Quevedo and the ICTP for their hospitality during final phase of work on this article. MD would also like to thank Markus Thoma for enlightening discussion on hard thermal loop resummation. 
This work was supported by the Gottfried Wilhelm Leibniz program of the Deutsche Forschungsgemeinschaft and the the Project of Knowledge Innovation
Program of the Chinese Academy of Sciences grant KJCX2.YW.W10.

\begin{appendix}
\section{The $\chi_i$ self-energy}\label{appendix}
In order to evaluate (\ref{OffShell}) we need $\Pi^-_{\p i}(p_0)$ for off-shell energies and arbitrary momenta.
The mass correction is dominated by the tadpole diagram shown in figure \ref{phi4}a) and can be estimated by (\ref{thermalmass}).
The self-energy $\Pi^-$ for $\chi_i$ is given by the diagram shown in figure \ref{phi4}a). It can be obtained from (\ref{risingsun}) with the replacement $h_i\rightarrow\lambda_i$.
This integral can be conveniently rewritten in the form
\begin{eqnarray}
\Pi^-_{\p i}(p_0)=2 i \big(\mathcal{D}^{[i]}_{\p}(p_0)+\mathcal{S}^{[i]}_\p(p_0)-\mathcal{D}^{[i]}_\p(-p_0)-\mathcal{S}^{[i]}_{\p}(-p_0)\big)
\end{eqnarray}
with
\begin{eqnarray}
\mathcal{D}^{[i]}_{\p}(p_0)&=&\pi\frac{\lambda_i^2}{24(2\pi)^5}\theta(p_0)
\int_{M_i}^{p_0-2M_i} d\Omega_{\k} \int_{M_i}^{\Omega^-} d\Omega_{\l}  F_d(\Omega_{\k},\Omega_{\l},p_0,A)\mathcal{I}(Z)\nonumber\\
\mathcal{S}^{[i]}_{\p}(p_0)&=&3\pi\frac{\lambda_i^2}{24(2\pi)^5}\theta(p_0)
\int_{M_i}^\infty d\Omega_{\k} \int_{\Omega^+}^\infty d\Omega_{\l} F_s(\Omega_{\k},\Omega_{\l},p_0,-A)\mathcal{I}(Z)\nonumber
\end{eqnarray}
with
\begin{eqnarray}
F_d(\Omega_{\k},\Omega_{\l},A)&=&
\big(1+f_B(\Omega_{\l})\big)
\big(1+f_B(\Omega_{\k})\big)
\big(1+f_B(A)\big)-
f_B(\Omega_{\l}) f_B(\Omega_{\k}) f_B(A)\nonumber\\
F_s(\Omega_{\k},\Omega_{\l},A)&=&
\big(1+f_B(\Omega_{\l})\big)
\big(1+f_B(\Omega_{\k})\big)
f_B(A)-
f_B(\Omega_{\l}) f_B(\Omega_{\k}) \big(1+f_B(A)\big)\nonumber\\
\mathcal{I}(Z)&=&\int_{-1}^{1}dx \int_{-1}^1 dy \frac{\theta(I(x,y,Z))}{\sqrt{I(x,y,Z)}}\nonumber\\
I(x,y,Z)&=&(1-x^2)(1-y^2)-(Z-xy)^2\nonumber
\end{eqnarray}
\begin{eqnarray}
A&=&p_0-\Omega_{\k}-\Omega_{\l} \ , \ B^2=\p^2+\k^2+\l^2+M_i^2\nonumber
\end{eqnarray}
\begin{eqnarray}
\Omega^\pm={\rm max}\big[p_0-\Omega_{\k}\pm M_i,M_i\big] \ , \
Z=\frac{ A^2 - B^2 + 2y |\p| |\k| + 2x |\p| |\l|}{2 |\k| |\l|}\nonumber
\end{eqnarray}
The variables $x$ nd $y$ are the cosines of the two nontrivial integration angles.
Note that the integration limits imply that $\mathcal{D}^{[i]}_{\p}(p_0)$ vanishes for $p_0<3M_i$, as expected from energy and momentum conservation in the decay and inverse decay processes $\chi_i\leftrightarrow \chi_i\chi_i\chi_i$.

\end{appendix}

\end{document}